\documentclass[A4]{svmult}

\usepackage{makeidx}         % allows index generation
\usepackage{graphicx}        % standard LaTeX graphics tool
\usepackage{multicol}        % used for the two-column index
\usepackage[bottom]{footmisc}% places footnotes at page bottom
\usepackage{algorithm,algorithmicx,algpseudocode}
\usepackage{graphicx}
\usepackage{subfigure}
\usepackage{amssymb}
\usepackage{multirow}
\usepackage{hyperref}
\usepackage{amsmath}
\usepackage{adjustbox}

\usepackage{tabularx}

\usepackage{array}
\newcommand\MyBox[2]{
  \fbox{\lower0.75cm
    \vbox to 1.7cm{\vfil
      \hbox to 1.7cm{\hfil\parbox{1.4cm}{#1\\#2}\hfil}
      \vfil}%
  }%
}

\makeindex             % used for the subject index
\newcommand{\bs}{\boldsymbol}
\newcommand{\mf}{\mathbf}

\newcommand{\tutr}{3$'$UTR}

\newcommand{\threeutr}{3$'$UTR~}

\newcommand{\wc}{Watson-Crick}

\newcommand{\etal}{\emph{et al.}~}

\usepackage{color}

\begin{document}

\title*{Unsupervised Learning in Genome Informatics}
%\author{Name of Author\inst{1}
%Name of Author\inst{2}}
\author{
Ka-Chun Wong\inst{1} 
Yue Li\inst{2}  
Zhaolei Zhang\inst{3}
}
\institute{
City University of Hong Kong
\texttt{kc.w@cityu.edu.hk}
\and
University of Toronto
\texttt{yueli@cs.toronto.edu}
\and
University of Toronto
\texttt{zhaolei.zhang@utoronto.ca}
}
%\and Name and Address of your Institute \texttt{name@email.address}}
\maketitle

With different genomes available, unsupervised learning algorithms are essential in learning genome-wide biological insights. Especially, the functional characterization of different genomes is essential for us to understand lives. In this book chapter, we review the state-of-the-art unsupervised learning algorithms for genome informatics from DNA to MicroRNA.

DNA (DeoxyriboNucleic Acid) is the basic component of genomes. A significant fraction of DNA regions (transcription factor binding sites) are bound by proteins (transcription factors) to regulate gene expression at different development stages in different tissues. To fully understand genetics, it is necessary of us to apply unsupervised learning algorithms to learn and infer those DNA regions. Here we review several unsupervised learning methods for deciphering the genome-wide patterns of those DNA regions.

MicroRNA (miRNA), a class of small endogenous non-coding RNA (RiboNucleic acid) species, regulate gene expression post-transcriptionally by forming imperfect base-pair with the target sites primarily at the 3$'$ untranslated regions of the messenger RNAs. Since the 1993 discovery of the first miRNA \emph{let-7} in worms, a vast amount of studies have been dedicated to functionally characterizing the functional impacts of miRNA in a network context to understand complex diseases such as cancer. Here we review several representative unsupervised learning frameworks on inferring miRNA regulatory network by exploiting the static sequence-based information pertinent to the prior knowledge of miRNA targeting and the dynamic information of miRNA activities implicated by the recently available large data compendia, which interrogate genome-wide expression profiles of miRNAs and/or mRNAs across various cell conditions.

\section{Introduction}

Since the 1990s, the whole genomes of a large number of species have been sequenced by their corresponding genome sequencing projects. In 1995, the first free-living organism \textit{Haemophilus influenzae} was sequenced by the Institute for Genomic Research \cite{pmid7542800}. In 1996, the first eukaryotic genome (\textit{Saccharomyces cerevisiase}) was completely sequenced \cite{yeastGenome}. In 2000, the first plant genome, \textit{Arabidopsis thaliana}, was also sequenced by Arabidopsis Genome Initiative \cite{plantGenome}. In 2004, the Human Genome Project (HGP) announced its completion \cite{pmid15496913}. Following the HGP, the Encyclopedia of DNA Elements (ENCODE) project was started, revealing massive functional putative elements on the human genome in 2011 \cite{pmid22955616}. The drastically decreasing cost of sequencing enables the 1000 Genomes Project to be carried out, resulting in an integrated map of genetic variation from 1,092 human genomes published in 2012 \cite{pmid23128226}. The massive genomic data generated by those projects impose an unforeseen challenge for large-scale data analysis at the scale of gigabytes or even terabytes. 

Computational methods are essential in analyzing the massive genomic data. They are collectively known as bioinformatics or computational biology; for instance, motif discovery \cite{compSurvey} helps us distinguish real signal subsequence patterns from background sequences.  Multiple sequence alignment \cite{blast} can be used to study the similarity between multiple sequences. Protein structure prediction \cite{proteinstructurepredictoin,wong2010protein} can be applied to predict the 3D tertiary structure from an amino acid sequence. Gene network inference \cite{geneNetwork} are the statistical methods to infer gene networks from correlated data (e.g. microarray data). Promoter prediction \cite{promoterPrediction} help us annotate the promoter regions on a genome. Phylogenetic tree inference \cite{phylogeneticInference} can be used to study the hierarchical evolution relationship between different species. Drug scheduling \cite{drugScheduling,wong2012evolutionary} can help solve the clinical scheduling problems in an effective manner. Although the precision of those computational methods is usually lower than the existing wet-lab technology, they can still serve as useful preprocessing tools to significantly narrow search spaces. Thus prioritized candidates can be selected for further validation by wet-lab experiments, saving time and funding. In particular, unsupervised learning methods are essential in analyzing the massive genomic data where the ground truth is limited for model training. Therefore, we describe and review several unsupervised learning methods for genome informatics in this chapter. 
%A discussion with future insights is provided at the end.

\section{Unsupervised Learning for DNA}

In human and other eukaryotes, gene expression is primarily regulated by the DNA binding of various modulatory transcription factors (TF) onto cis-regulatory DNA elements near genes. Binding of different combinations of TFs may result in a gene being expressed in different tissues or at different developmental stages. To fully understand a gene's function, it is essential to identify the TFs that regulate the gene and the corresponding TF binding sites (TFBS). Traditionally, these regulatory sites were determined by labor-intensive experiments such as DNA footprinting or gel-shift assays. Various computational approaches have been developed to predict TF binding sites \emph{in silico}. Detailed comparisons can be found in the survey by Tompa et al.  \cite{Tompa}. TFBS are relatively short (10-20 bp) and highly degenerate sequence motifs, which makes their effective identification a computationally challenging task. A number of high-throughput experimental technologies were developed recently to determine protein-DNA binding such as protein binding microarray (PBM) \cite{pmid16998473}, chromatin immunoprecipitation (ChIP) followed by microarray or sequencing (ChIP-Chip or ChIP-Seq) \cite{chipchip,chipseq}, microfluidic affinity analysis \cite{pmid20802496}, and protein microarray assays \cite{pmid19879846,pmid16785442} .

On the other hand, it is expensive and laborious to experimentally identify TF-TFBS sequence pairs, for example, using DNA footprinting \cite{Footprint}, gel electrophoresis \cite{Gel_shift}, and SELEX \cite{pmid2200121}. The technology of Chromatin immunoprecipitation (ChIP) \cite{chipchip,chipseq} measures the binding of a particular TF to the nucleotide sequences of co-regulated genes on a genome-wide scale \emph{in vivo}, but at low resolution. Further processing are needed to extract precise TFBSs \cite{pmid12101404}. Protein Binding Microarray (PBM) was developed to measure the binding preference of a protein to a complete set of k-mers  \emph{in vitro} \cite{pmid16998473}. The PBM data resolution is unprecedentedly high, comparing with the other traditional techniques. The DNA k-mer binding specificities of proteins can even be determined in a single day. It has also been shown to be largely consistent with those generated by  \emph{in vivo} genome-wide location analysis (ChIP-chip and ChIP-seq) \cite{pmid16998473}.

To store and organize the precious data, databases have been created. TRANSFAC is one of the largest databases for regulatory elements including TFs, TFBSs, weight matrices of the TFBSs, and regulated genes \cite{TRANSFAC06}. JASPAR is a comprehensive collection of TF DNA-binding preferences \cite{JASPAR2010}. Other annotation databases are also available (e.g. Pfam \cite{Pfam2004}, UniProbe \cite{pmid21037262}, ScerTF \cite{ScerTF}, FlyTF \cite{FlyTF}, YeTFaSCo \cite{pmid22102575}, and TFcat \cite{TFcat}). Notably, with the open-source and open-access atmosphere wide-spreads on the Internet in recent years, a database called ORegAnno appeared in 2008 \cite{ORegAnno}. It is an open-access community-driven database and literature curation system for regulatory annotation. The ENCODE consortium has also released a considerable amount of ChIP-Seq data for different DNA-binding proteins \cite{pmid22955616}.

In contrast, unfortunately,  it is still difficult and time-consuming to extract the high-resolution 3D protein-DNA (e.g. TF-TFBS ) complex structures with X-Ray Crystallography \cite{xray} or Nuclear Magnetic Resonance (NMR) spectroscopic analysis \cite{NMR}. As a result, there is strong motivation to have unsupervised learningl methods based on existing abundant sequence data, to provide testable candidates with high confidence to guide and accelerate wet-lab experiments. Thus unsupervised learning methods are proposed to provide insights into the DNA binding specificities of transcription factors from the existing abundant sequence data.

\subsection{DNA Motif Discovery and Search}
Transcription Factor Binding Sites (TFBSs) are represented in DNA motif models to capture its sequence degeneracy \cite{Wong:2011:GLP:2093590.2093604}. They are described in the following sections.

\subsubsection{Representation (DNA Motif Model)}
There are several motif models proposed. For example, consensus string representation, a set of motif instance strings, count matrix, position frequency matrix (PFM), and position weight matrix (PWM). Among them, the most popular motif models are the matrix ones. They are the count matrix, PFM, and PWM. In particular, the most common motif model is the zero-order PWM which has been shown to be related to the average protein-DNA binding energy in the experimental and statistical mechanics study \cite{pmid3612791}. Nonetheless, it assumes independence between different motif positions. A recent attempt has been made to generalize PWM but the indel operations between different nucleotide positions are still challenging \cite{alllevelcomplexity}. Although the column dependence and indel operations could be modeled by Hidden Markov Model (HMM) simultaneously, the number of training parameters is increased quadratically. There is a dilemma between accuracy and model complexity.

\paragraph{Count Matrix}

\begin{figure*}[h]
\centering
\includegraphics[width=0.5\textwidth]{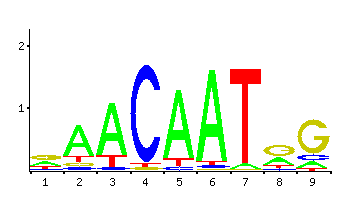}
\caption{DNA Motif Sequence Logo for SOX9 (JASPAR ID:MA0077.1 and UniProt ID:P48436). The vertical axis measures the information content, while the horizontal axis denotes the positions. The consensus string is DAACAATRG, following the standard IUPAC nucleotide code.}
\label{fig:logo}
\end{figure*}

The count matrix representation is the \textit{de facto} standard adopted in databases. In the count matrix representation, a DNA motif of width $w$ is represented as a $4$-by-$w$ matrix $C$. The $j$th column of $C$ corresponds to the $j$th position of the motif, whereas the $i$th row of $C$ correspond to the $i$th biological character. In the context of DNA sequence, we have 4 characters $\{$A,C,G,T$\}$. $C_{ij}$ is the occurring frequency of the $i$ biological character at the $j$th position. For example, the count matrix $C_{sox9}$ of the SOX9 protein (JASPAR ID:MA0077.1 and UniProt ID:P48436) is tabulated in the following matrix form. The motif width is 9 so we have a $4\times9$ matrix here. The corresponding sequence logo is also depicted in Figure \ref{fig:logo}. 
\[
C_{sox9} = \bordermatrix{
~ & 1 & 2 & 3 & 4 & 5 & 6 & 7 & 8 & 9 \cr
A & 24 & 54 & 59 & 0 & 65 & 71 & 4 & 24 & 9 \cr
C & 7 & 6 & 4 & 72 & 4 & 2 & 0 & 6 & 9 \cr
G & 31 & 7 & 0 & 2 & 0 & 1 & 1 & 38 & 55 \cr
T & 14 & 9 & 13 & 2 & 7 & 2 & 71 & 8 & 3 \cr
}
\]

Notably,  adenine has not been found at the 4th position, resulting in a zero value. It leads to an interesting scenario. Is adenine really not found at that position or the sample size (the number of binding sites we have found so far for SOX9) is too small that the sites having adenine at that position have not been verified experimentally yet ? To circumvent the problem, people usually add pseudo counts to the matrix which is justified from the use of prior probability in statistics \cite{citeulike:163532,pmid19106141}. Such techniques are also found in natural language computing and machine learning.

\paragraph{Position Frequency Matrix (PFM)}

In practice, a count matrix is usually converted to PFM, and thus a zero-order PWM for scanning a long sequence. The dimension and layout of count matrix is exactly the same as those of the corresponding PFM and zero-order PWM. Their main difference is the element type. For count matrix, each element is simply a count. For PFM, each element is a Maximum Likelihood Estimate (MLE) parameter.  For zero-order PWM, each element is a weight. 

To derive a PFM $F$ from a count matrix, $C$, maximum likelihood estimation (MLE) is used \cite{mosesSurvey}. Mathematically, we aim at maximizing the likelihood function $L(C) =  P(C|F) = \prod_{i=1}^{w} \prod_{j=1}^{4} F_{ji}^{C_{ji}}$. In addition, we impose a parameter normalization constraint $\sum_{j=1}^{4}F_{ji} = 1$ for each $ith$ position. It is added to the likelihood function with a Lagrange multiplier $\lambda_i$, resulting in a new log likelihood function:  \[ln L'(C) = \sum_{i=1}^{w} \sum_{j=1}^{4} C_{ji} log (F_{ji})+\sum_{i=1}^{w} \lambda_i(\sum_{j=1}^{4} F_{ji}-1)\]By taking its partial derivatives to zero, it has been shown that $F_{ji} = \frac{C_{ji}}{\sum_{j=1}^{4}C_{ji}}$. The MLE parameter definition is quite intuitive. It is simply the occurring fraction of a nucleotide at the same position. For example, given the previous SOX9 count matrix $C_{sox9}$, we can convert it to a PFM $F_{sox9}$ as follows:
\[
F_{sox9} = \bordermatrix{
~ & 1 & 2 & 3 & 4 & 5 & 6 & 7 & 8 & 9 \cr
%A & 24 & 54 & 59 & 0 & 65 & 71 & 4 & 24 & 9 \cr
%C & 7 & 6 & 4 & 72 & 4 & 2 & 0 & 6 & 9 \cr
%G & 31 & 7 & 0 & 2 & 0 & 1 & 1 & 38 & 55 \cr
%T & 14 & 9 & 13 & 2 & 7 & 2 & 71 & 8 & 3 \cr
%A&0.32 &0.71 &0.78 &0.00 &0.86 &0.93 &0.05 &0.32 &0.12  \cr
%C&0.09 &0.08 &0.05 &0.94 &0.05 &0.03 &0.00 &0.08 &0.12  \cr
%G&0.41 &0.09 &0.00 &0.03 &0.00 &0.01 &0.01 &0.50 &0.72  \cr
%T&0.18 &0.12 &0.17 &0.03 &0.09 &0.03 &0.94 &0.10 &0.04  \cr
A&0.31 &0.68 &0.75 &0.01 &0.83 &0.89 &0.06 &0.31 &0.13 \cr
C&0.10 &0.09 &0.06 &0.91 &0.06 &0.04 &0.01 &0.09 &0.13 \cr
G&0.40 &0.10 &0.01 &0.04 &0.01 &0.03 &0.03 &0.49 &0.69 \cr
T&0.19 &0.13 &0.18 &0.04 &0.10 &0.04 &0.90 &0.11 &0.05 \cr
}
\]
where a pseudocount = 1 is added to each element of $C_{sox9}$.

We can observe that the most invariant positions of the SOX9 motif are the 4th and 6th position. At the 4th position, cytosines have been found most of the times while guanines and thymines have just been found few times. 

\paragraph{Position Weight Matrix (PWM)}

To scan a long sequence for motif matches using a PFM $F$, we need to derive a PWM $M$ first so that the background distribution can be taken into account. As aforementioned, each element of PWM is a weight. Each weight can be viewed as a preference score. In practice, it is usually defined as the log likelihood ratio between the motif model and background model. Mathematically, $M_{ji} = log(\frac{F_{ji}}{B_j})$ where $B_j$ is the occurring fraction of the $j$th nucleotide in all the background sequences such that, given a subsequence $a$ of the same width as the PWM (width $w$), we can compute a score $S(a)$ by summation only:
\begin{eqnarray}
S(a) & = & log\frac{P(a|F)}{P(a|Background)} \\
& = & log\frac{\prod_{i=1}^{w} \prod_{j=1}^{4} F_{ji}^{[a_i=j] }}{\prod_{i=1}^{w} \prod_{j=1}^{4} B_j^{[a_i=j] }} \\
& = & log\prod_{i=1}^{w} \prod_{j=1}^{4}  (\frac{F_{ji}^{[a_i=j] }}{B_j^{[a_i=j] }}) \\
& = & \sum_{i=1}^{w} \sum_{j=1}^{4} [a_i=j] log(\frac{F_{ji}}{B_j}) \\
& = & \sum_{i=1}^{w} \sum_{j=1}^{4} [a_i=j]M_{ji}
\end{eqnarray}
where $a_i$ is the numeric index for the nucleotide at $i$th position of the subsequence $a$. For example, given the previous SOX9 PFM $F_{sox9}$, we can convert it to a PWM $M_{sox9}$ as follows:
\scriptsize
\[
M_{sox9} = \bordermatrix{
~ & 1 & 2 & 3 & 4 & 5 & 6 & 7 & 8 & 9 \cr
%A & 24 & 54 & 59 & 0 & 65 & 71 & 4 & 24 & 9 \cr
%C & 7 & 6 & 4 & 72 & 4 & 2 & 0 & 6 & 9 \cr
%G & 31 & 7 & 0 & 2 & 0 & 1 & 1 & 38 & 55 \cr
%T & 14 & 9 & 13 & 2 & 7 & 2 & 71 & 8 & 3 \cr
%A&0.32 &0.71 &0.78 &0.00 &0.86 &0.93 &0.05 &0.32 &0.12  \cr
%C&0.09 &0.08 &0.05 &0.94 &0.05 &0.03 &0.00 &0.08 &0.12  \cr
%G&0.41 &0.09 &0.00 &0.03 &0.00 &0.01 &0.01 &0.50 &0.72  \cr
%T&0.18 &0.12 &0.17 &0.03 &0.09 &0.03 &0.94 &0.10 &0.04  \cr
A&0.32 &1.46 &1.58 &-4.32 &1.72 &1.85 &-2.00 &0.32 &-1.00 \cr
C&-1.32 &-1.51 &-2.00 &1.87 &-2.00 &-2.74 &-4.32 &-1.51 &-1.00 \cr
G&0.68 &-1.32 &-4.32 &-2.74 &-4.32 &-3.32 &-3.32 &0.96 &1.49 \cr
T&-0.42 &-1.00 &-0.51 &-2.74 &-1.32 &-2.74 &1.85 &-1.15 &-2.32 \cr
}
\]
\normalsize
where we have assumed that the background distribution is uniform (i.e. $B_j=0.25$ for $1\leq j \leq 4$) for illustration purposes. 
%A pseudocount of 1 is added to each element of $C_{sox9}$.

\subsubsection{Learning (Motif Discovery)}
In general, motif discovery aims at building motif models (e.g. PFM) from related sequences. Nonetheless, there is a variety of motif discovery methods in different biological settings. From a computing perspective, they can be classified into several paradigms by its input data types: 

\begin{enumerate}
\item A Set of Sequences

\item A Set of Sequences with Quantitative Measurements

\item A Set of Orthologous Sequences
\end{enumerate}

\paragraph{Motif Discovery for A Set of Sequences}
The most classical one is \textit{de novo} motif discovery which just takes a set of sequences as the inputs. The set of sequences is extracted such that a common transcription factor is believed to bind to them, assuming that motif models (e.g. consensus substrings) can be found from those sequences. For example, the promoter and enhancer sequences of the genes co-regulated by a common transcription factor or the sequence regions around the next generation parallel sequencing peaks called for a common transcription factor. Theoretically, Zia and Moses have proved a theoretical upper bound on the p-value which at least one motif with a specific information content occur by chance from background distribution (false positive) for the one-occurrence per sequence motif discovery problem \cite{pmid22738169}.

Chan et al. applied evolutionary computation techniques to the problem \cite{pmid18065426,kcwong:EASE}. Hughes et al. proposed a Gibbs sampling algorithm called AlignACE, to sample and evaluate different possible motif models using a priori log likelihood scores \cite{pmid10698627}. Workman et al. have proposed a machine learning approach using artificial neural networks (ANN-Spec) \cite{pmid10902194}. Hertz et al. utilized the maximal information content principle to greedily search for a set of candidate sequences for building motif models (Consensus) \cite{pmid10487864}. Frith et al. have adopted simulated annealing approaches to perform multiple local alignment for motif model building (GLAM) \cite{pmid14704356}. Ao et al. have used expectation maximization to determine DNA motif position weight matrices (Improbizer) \cite{pmid15375261}. Bailey et al. have proposed MEME to optimize the expected value of a statistic related to the information content of motif models \cite{pmid7584439}. A parameter enumeration pipeline wrapping MEME (MUSI) was proposed for elucidating multiple specificity binding modes \cite{pmid22210894}. Eskin et al. employed a tree data structure to find composite weak motifs (MITRA) \cite{pmid12169566}. Thijs et al. have further improved the classic Gibbs sampling method and called it MotifSampler \cite{pmid11751219}. Van Helden et al. have proposed a counting algorithm to detect statistically significant motifs \cite{pmid9719638}. Regnier and Denise have proposed an exhaustive search algorithm (QuickScore) \cite{regnier04:_rare}. Favorov et al. have utilized Markov Chain Monte Carlo to solve the problem in a Bayesian manner (SeSiMCMC) \cite{pmid15728117}. Pavesi  et al. have proposed an exhaustively enumerated and consensus-based method (Weeder) \cite{pmid15215380}. Another exhaustive search algorithm to optimize z-scores (YMF) has been proposed by Sinha and Tompa \cite{pmid12824371}. 

Although different statistical techniques have been developed for the motif discovery problem, most of the existing methods aim at building motif models in the form of either a set of strings or a zero-order PWM. Nonetheless, it is well known that nucleotide dependencies and indel operations exist within some TFBSs (a.k.a. motifs) \cite{pmid17308339,pmid18184687}. It is desirable to develop new methods which can capture and model such information.

\paragraph{Motif Discovery for A Set of Sequences with Quantitative Affinity Measurements}
%With the advancement of Chip and PBM technology, each sequence can be associated with a binding intensity value. Although we can still apply the previous methods with an ad hoc threshold, it is more appropriate to develop new types of algorithms to take those quantitative measurements into account. In light of that, Seed and Wobble has been proposed as a seed-based approach using rank statistics \cite{pmid16998473}. RankMotif++ was proposed to maximize the log likelihood of their probabilistic model of binding preferences \cite{pmid17646348}. MatrixREDUCE is proposed to  perform forward variable selections to minimize the sum of squared deviations \cite{pmid16317069}. MDScan is proposed to combine two search strategies together, namely word enumeration and position-specific weight matrix updating \cite{pmid12101404}. PREGO is proposed to maximize the Spearman rank correlation between the predicted and the actual binding intensities \cite{pmid16809671}. 

It has been pointed out that a fundamental bottleneck in TFBS identification is the lack of quantitative binding affinity data for a large portion of transcription factors. The advancement of new high-throughput technologies such as ChIP-Chip, ChIP-Seq, and Protein Binding Microarray (PBM) has made it possible to determine the binding affinity of these TFs (i.e. each sequence can be associated with a binding intensity value) \cite{pmid19443739}. In particular, the PBM technology can enable us to enumerate all the possible k-mers, providing an unprecedentedly high resolution binding site affinity landscape for each TF. In light of this deluge of quantitative affinity data, robust probabilistic methods were developed to take into account those quantitative affinity data. Seed and Wobble has been proposed as a seed-based approach using rank statistics \cite{pmid16998473}. RankMotif++ was proposed to maximize the log likelihood of their probabilistic model of binding preferences \cite{pmid17646348}. 
%Other methods were also proposed \cite{pmid16317069,pmid12101404,pmid16809671}
MatrixREDUCE was proposed to  perform forward variable selections to minimize the sum of squared deviations \cite{pmid16317069}. MDScan was proposed to combine two search strategies together, namely word enumeration and position-specific weight matrix updating \cite{pmid12101404}. PREGO was proposed to maximize the Spearman rank correlation between the predicted and the actual binding intensities \cite{pmid16809671}. Notably, Wong et al. have proposed and developed a hidden Markov model approach to learn the dependence between adjacent nucleotide positions rigorously; they also show that their method (kmerHMM) can deduce multiple binding modes for a given TF \cite{pmid23814189}.

Note that this paradigm is a generalization from the motif discovery for a set of sequences with binary measurements. For example, SeedSearch \cite{Barash:2001:SHA:645906.673098} and DME \cite{pmid15668401}. In other words, it also includes the discriminative motif discovery method in which a set of motif-containing sequences and a set of background sequences are given as the input since we can assign a value of 1 to each motif-containing sequence and 0 to each background sequence. 

\paragraph{Motif Discovery for A Set of Orthologous Sequences}
%In recent years, genome sequencing projects around the world have successfully sequenced the whole genomes of different species . In 1995, the first free-living organism \textit{Haemophilus influenzae} was sequenced by The Institute for Genomic Research\cite{pmid7542800}. In 1996, the first eukaryotic genome was completely sequenced. It was the model eukaryote species, \textit{Saccharomyces cerevisiase}\cite{yeastGenome}. In 2000, the first plant genome, \textit{Arabidopsis thaliana}, was also sequenced by Arabidopsis Genome Initiative\cite{plantGenome}. Finally in 2003, the Human Genome Project (HGP) also announced its completion \cite{citeulike:197238}. 
%To make better use of the genomes, comparative genome information can be incorporated into motif discovery.
It is generally acknowledged that by comparing evolutionarily related DNA or protein sequences, functionally important sequences or motifs can be revealed by such comparison. 
%People exploited the fact that mutations on motifs could affect the fitness of a individual. The individuals with deleteriously mutated motifs should have been removed during the positive selection in nature. 
%Deleterious mutations will be removed from population by negative selection. 
A functional motif is assumed to be more conserved across different species than the background sequences \cite{15534694}. By incorporating the evolutionary conservation with sequence-specific DNA binding affinity, different methods have been proposed. Moses et al. have proposed an extension of MEME to take the sequence evolution into account probabilistically \cite{pmid14992514}. Kellis et al. have proposed a spaced hexamer enumeration approach to identify conserved motifs \cite{pmid15285895}. FootPrinter has been proposed as a substring-parsimony based approach using dynamic programming to find statistically enriched motif substrings \cite{pmid12015878}. A Gibbs sampling approach named PhyloGibbs has also been proposed \cite{pmid16477324}.

\subsubsection{Prediction (Motif Search)}
After a motif model has been found, it is always desirable to apply it to search for motif instances over a given sequence (e.g. ChIP-Seq peak sequences). Some basic search methods have been developed to search motif instances over a sequence. Nonetheless, those methods do not have sufficient motif model complexity to distinguish false positives from true positives over a long sequence (e.g. 100k bp) \cite{pmid15131651}. To cope with that, some improvements have been made. In general, most of them utilize the biological information beyond the motif sequence specificity to augment the motif model complexity insufficiency. In particular, multiple motif information and evolutionary conservation have been readily adopted to improve the discovery accuracy \cite{mosesSurvey}.

\paragraph{Basic Searches}

\paragraph{Likelihood Ratio}
Given a sequence $b_1b_2b_3...b_l$ of length $l$ and a PWM $M$ of the motif $x$ of width $w$, we scan $b_1b_2b_3...b_l$ with a window of width $w$ such that the subsequences which likelihood ratio score is higher than a pre-specified threshold are considered the instances (hits) of the motif $x$. Mathematically, a subsequence $b_{k+1}b_{k+2}...b_{k+w}$ is considered as a motif instance (hit) if and only if the following condition is satisfied:
\[
S_{x}(b_{k+1}b_{k+2}...b_{k+w}) = \sum_{i=1}^{w} \sum_{j=1}^{4} I(b_{k+i}=n_j)M_{ji} > threshold
\]
where $n_j$ is the $j$th nucleotide among \{A,C,G,T\} and $I(...)$ is the Iverson bracket.
Nonetheless, different motifs may have different likelihood score distributions. It is difficult to set a single and fixed threshold which can work for all the motifs. To solve the problem, one can normalize the score to the interval [0,1] based on the maximal and minimal scores as follows:
\[
S'(b_{k+1}b_{k+2}...b_{k+w}) = \frac{S_x(b_kb_{k+1}b_{k+2}...b_{n}) -\min\limits_{seq} S_x(seq)}{\max\limits_{seq} S_x(seq)-\min\limits_{seq} S_x(seq)}
\]

\paragraph{Posterior Ratio}
If we know the prior probability of motif occurrence $\pi$, it can also be incorporated into the scoring function in a posterior manner \cite{mosesSurvey}. Mathematically, given a sequence $a$,  motif model (including PFM $F$ and PWM $M$),  and background distribution $B$, we can compute the posterior ratio as follows:
\begin{eqnarray}
S''(a) & = & log\frac{P(F|a)}{P(B|a)} \\
& = & log\frac{\frac{P(a|F)P(F)}{P(a)}}{\frac{P(a|B)P(B)}{P(a)}} \\
& = & log\frac{P(a|F)P(F)}{P(a|B)P(B)} \\
& = & log\frac{\prod_{i=1}^{w} \prod_{j=1}^{4}  F_{ji}^{[a_i=j]}\pi}{\prod_{i=1}^{w} \prod_{j=1}^{4} B_j^{[a_i=j]}(1-\pi)} \\
& = & log\prod_{i=1}^{w} \prod_{j=1}^{4} (\frac{F_{ji}}{B_j})^{ [a_i=j]}(\frac{\pi}{1-\pi}) \\
& = & S(a) + log(\frac{\pi}{1-\pi}) 
\end{eqnarray}
It can be observed that the posterior ratio can be computed from the likelihood ratio by simply adding the logarithm of the prior probability ratio.

\paragraph{P-value}
Given the previous scoring functions, it is not easy to set a threshold since they are just ratios. For example, if $S(a)>0$ in the above example, it just means the likelihood that the sequence $a$ is generated by the motif model is higher than the background and vice versa. To justify it in a meaningful way, P-value distribution can be calculated from a motif model. Given a motif PWM $M$ of width $w$, an exhaustive search can be applied to traverse all the possible sequences of width $w$. Nonetheless, it takes $4^w$ time complexity for the DNA alphabet $\{A,C,G,T\}$. Interestingly, if the PWM $M$ is of zero-order, we can exploit the column independence assumption and apply dynamic programming to calculate the exact P-value distribution in $4w$ time complexity \cite{pmid2720468}. In practice, the empirical P-value distribution may also be used.

Nonetheless, the specificity of a PWM of width $w$ is still not high if it is applied to a very long sequence of length $L$. Mathematically, even if we just assign the best match as the hit, $\frac{L-w+1}{4^w}$ hits are still expected (e.g. If $L=10000$ and $w=6$, $2.44$ hits are expected), assuming that the sequence is uniform in background nucleotide distribution. To solve the problem, people have spent efforts on incorporating more biological information to improve the motif search.

\paragraph{Incorporating Multiple Motif Information}
To improve motif search, multiple motif information can be incorporated. Multiple motif sites are usually clustered together, resulting in higher signal-to-noise ratios which can be easier to be detected than alone. If multiple sites of the same motif are clustered together within a short distance, it is called homotypic clustering \cite{pmid12670999}. On the other hand, if multiple sites of different motifs are clustered together within a short distance, it is called heterotypic clustering.

To exploit the additional clustering signals beyond sequence specificity, MAST was proposed to multiply the P-values of multiple motif matches (hits) together, which has demonstrated superior performance in sequence homology search than the other two methods proposed in the same study \cite{pmid9672829}. CIS-ANALYST was proposed as a sliding window approach to predict the windows which have at least $min\_sites$ motif matches (hits) with pvalues $<site\_p$ \cite{pmid11805330}. Sinha et al. have proposed a probabilistic model, Stubb, to efficiently detect clusters of binding sites (i.e. cis-regulatory modules) over genomic scales using maximum likelihood estimation \cite{pmid16845069}. To determine the window size parameter, a window size adjustment procedure has been used in ClusterBuster to find clusters of motif matches \cite{pmid12824389}. Segal et al. have also derived an expectation maximization algorithm to model the clusters of motif matches as probabilistic graphical models \cite{pmid12855470}. Recently, Hermann et al. have proposed an integrative system (i-cisTarget) to combine the high-throughput next generation sequencing data with motif matches to provide accurate motif cluster search \cite{pmid22718975}. Notably, Zhou and Wong have shown that it is possible to search for clusters of motifs in a \textit{de novo} way (i.e. without any given motif model and information) \cite{pmid15297614}. 

\paragraph{Incorporating Evolutionary Conservation}
Another approach to improve motif search is to incorporate evolutionary conservation. The rationale behind that is similar to that behind phylogenetic motif discovery which we have described in a previous section. Deleterious mutations will be removed from population by negative selection. 
%If an individual is shown to survive and gets sequenced, it means the individual is fit and functional. 
To make use of that fact, we could imply that a true motif match should be more conserved across closely related species (For example, chimpanzee and mouse) than background sequences \cite{15534694}. For instance, a windowing approach with several thresholds for motif matches and conservation, ConSite, was proposed by Sandelin et al. \cite{pmid15215389}. Nonetheless, it is limited to pair-wise analysis. rVISTA is a similar approach \cite{pmid16888351} using the Match program for motif matching in TRANSFAC \cite{TRANSFAC06}. Bayesian Branch Length Score (BBLS) was proposed as a evolutionary conservation score without relying on any multiple sequence alignment \cite{pmid19017655}. A parsimonious method for finding statistically significant k-mers with dynamic programming was proposed (FoorPrinter) \cite{pmid11997340}. Notably, Moses et al. proposed a comprehensive probabilistic model to search for motif instances with efficient p-value estimation (MONKEY) \cite{pmid15575972}. 

To search for novel motif instances, there are programs aimed at searching for motif matches without any given motif model and information. For instance, Ovcharenko et al. have used likelihood ratio tests to distinguish conserved regions from the background \cite{pmid15173121}. Siepel et al. have demonstrated an approach in identifying conserved regions using hidden Markov models called PhastCons \cite{pmid16024819}. 

\paragraph{Incorporating Both Approaches}
Both motif clustering information and evolutionary conservation were demonstrated beneficial to motif search. Since they are independent of each other, it is straightforward to combine them. Philippakis et al. have proposed a method to combine both types of information (i.e. motif clustering and evolutionary conservation), achieving good performance on experimentally verified datasets \cite{pmid15759656}. MONKEY has been extended by Warner et al. to exploit the motif clustering information to predict motif clusters (PhylCRM) \cite{pmid18311145}. It has been reported that the misalignment errors of the input reference sequences from other species could affect the quality of phylogenetic footprinting for motif search. Thus statistical alignments have been used to assist motif search in EMMA \cite{pmid19293946}. Stub has also been extended to StubMS to take multiple species conservation into account using a HMM phylogenetic model \cite{pmid12855472}. Notably, a unified probabilistic framework which integrates multiple sequence alignment with binding site predictions, MORPH, was proposed by the same group \cite{pmid17997594}. Its effectiveness has been demonstrated and verified in an independent comparison study \cite{pmid21152003}.

\subsection{Genome-wide DNA Binding Pattern Discovery}

Chromatin immunoprecipitation (ChIP) followed by high-throughput sequencing (ChIP-Seq) measures the genome-wide occupancy of transcription factors \textit{in vivo}. In a typical ChIP-Seq study, the first step is to call the peaks, i.e. determining the precise location in the genome where the TF binds. A number of peak calling tools have been developed; for instance, model-based analysis of ChIP-Seq data (MACS) was proposed to model the shift size of ChIP-Seq tags and local biases to improve its peak-calling accuracy \cite{pmid18798982}. Spp is another method with a strong focus on background signal correction \cite{pmid19029915}. PeakSeq is a two-pass strategy method. The first pass accounts for the sequence mappability while the second pass is to filter out statistically insignificant regions comparing to controls \cite{pmid19122651}. CisGenome refines peak boundaries and uses a conditional binomial model to identify peak regions \cite{pmid18978777}. However, recent benchmark studies suggest that their predicted peaks are distinct from each other \cite{pmid20628599,pmid20017957}.

Different combinations of DNA-binding protein occupancies may result in a gene being expressed in different tissues or at different developmental stages. To fully understand a gene's function, it is essential to develop unsupervised learning models on multiple ChIP-Seq profiles to decipher the combinatorial regulatory mechanisms by multiple transcription factors. 

Since multiple transcription factors often work in cis regulatory modules to confer complex gene regulatory programs, it is necessary to develop models on multiple ChIP-Seq datasets to decipher the combinatorial DNA-binding mechanism. 
In the following, we briefly review some of the previous works in this area. 
Gerstein et al. used pair-wise peak overlapping patterns to construct a human regulatory network \cite{pmid22955619}. Xie et al. proposed self organizing map methods to visualize the co-localization of DNA-binding proteins \cite{pmid24243024}. Giannopoulou et al. proposed a non-negative matrix factorization to elucidate the clustering of DNA-binding proteins \cite{pmid23554462}. Zeng and colleagues proposed jMOSAiCS to discover histone modification patterns across multiple ChIP-Seq datasets \cite{pmid23844871}. 
Ferguson et al. have described a hierarchical Bayes approach to integrate multiple ChIP-Seq libraries to improve DNA binding event predictions. 
In particular, they have applied the method to histone ChIP-Seq libraries and predicted the gene locations associated with the expected pathways \cite{ferguson2012new}. 
Mahony et al. also proposed a mixture model (MultiGPS) to detect differential binding enrichment of a DNA-binding protein in different cell lines, which can improve the protein's DNA binding location predictions (i.e. Cdx2 protein in their study) \cite{mahony2014integrated}. On the other hand, Chen et al. proposed a statistical framework (MM-ChIP) based on MACS to perform an integrative analysis of multiple ChIP datasets to predict ChIP-enriched regions with known motifs for a given DNA-binding protein (i.e. ER and CTCF proteins in their study) \cite{chen2011mm}. On the other hand, Ji et al. proposed a differential principal component analysis method on ChIP-Seq to perform unsupervised pattern discovery and statistical inference to identify differential protein-DNA interactions between two biological conditions \cite{ji2013differential}.
Guo et al. described a generative probabilistic model (GEM) for high resolution DNA binding site discovery from ChIP data \cite{guo2012high}. Interestingly, that model combines ChIP signals and DNA motif discovery together to achieve precise predictions of the DNA binding locations of a DNA-binding protein. The authors have further demonstrated how GEM can be applied to reveal spatially constrained transcription factor binding site pairs on a genome.

Despite the success of the methods described above, to fully understand a gene's function, it is essential to develop probabilistic models on multiple ChIP-Seq profiles to decipher the genome-wide combinatorial patterns of DNA-binding protein occupancy. 
Unfortunately, the majority of the previous work usually focused on large-scale clustering of called peaks, which is an intuitive and straightforward approach. However such approaches have two limitations, as (i) peak-calling ignores the contributions from weak bindings of TFs, and (ii) pair-wise analysis ignores the complex combinatorial binding pattern among the TFs. Thus an unsupervised learning model called SignalSpider has been proposed to directly analyze multiple normalized ChIP-Seq signal profiles on all the promoter and enhancer regions quantitatively so that weak bindings can be taken into account \cite{wong2014signalspider,pmid22955978}. Especially, its computational complexity has been carefully designed to scale with the increasing ChIP-Seq data (i.e. linear complexity). With such a linear complexity, the method (SignalSpider) has been successfully applied to more than 100 ChIP-Seq profiles in an integrated way, revealing different genome-wide DNA-binding modules across the entire human genome (hg19) \cite{wong2014signalspider}.

%%%%%%%% Unsupervised Learning for inferring microRNA regulatory network %%%%%%%%

\section{Unsupervised Learning for inferring microRNA regulatory network}
\begin{figure*}[!t]
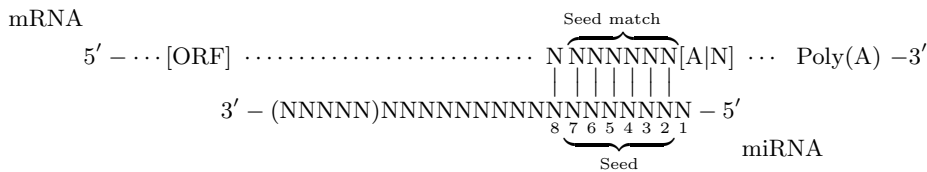

\begin{center}
\setlength{\jot}{-0.02in}
\begin{align*}
\raisebox{.20in}{mRNA}\text{5$'-\cdots$\text{[ORF]}$~\cdots\cdots\cdots\cdots\cdots\cdots\cdots\cdots\cdots~$}&\overset{\text{Seed match}}{\underset{\big|}{\text{N}}\overbrace{\underset{\big|}{\text{N}}\underset{\big|}{\text{N}}\underset{\big|}{\text{N}}\underset{\big|}{\text{N}}\underset{\big|}{\text{N}}\underset{\big|}{\text{N}}}}\text{[A$|$N]$~\cdots~$ Poly(A) $-3'$}\\
3'-\text{(NNNNN)NNNNNNNNN}&\underset{8}{\text{N}}\underset{\text{Seed}}{\underbrace{\underset{7}{\text{N}}\underset{6}{\text{N}}\underset{5}{\text{N}}\underset{4}{\text{N}}\underset{3}{\text{N}}\underset{2}{\text{N}}}}\underset{1}{\text{N}}-5'\raisebox{-.20in}{miRNA}
\end{align*}
\end{center}
\caption[Canonical miRNA Watson-Crick base pairing to \threeutr target site.]{Canonical miRNA Watson-Crick base pairing to the \threeutr of the mRNA target site. The most critical region is a 6mer site termed as the ``seed" occurs at the 2-7 position of the 5$'$ end of the miRNA \cite{Lewis:2003uz}. Three other variations centring at the 6mer seed are also known to be (more) conserved: 7mer-m8 site, a seed match +  a \wc~ match to miRNA nucleotide 8; 7mer-t1A site, a seed match + a downstream A in the \tutr; 8mer, a seed match + both m8 and t1A. The site efficacy has also been proposed in the order of 8mer $>$ 7mer-m8 $>$ 7mer- A1 $>$ 6mer \cite{Friedman:2009km,Grimson:2007cy}. The abbreviations are: ORF, open reading frame; (NNNNN), the additional nucleotides to the shortest 19 nt miRNA; [A$|$N], A or other nucleotides; Poly(A), polyadenylated tail.}
\label{background:fig:seedmatch}
\end{figure*}

While transcription factors (TFs) are the major transcriptional regulator proteins, microRNA (miRNA), a small $\sim$22 nucleotide noncoding RNA species, has been shown to play a crucial role in post-transcriptional and/or translational regulation \cite{Bartel:2009fh}. Since the 1993 discovery of the first miRNA \emph{let-7} in worms, a vast amount of studies have been dedicated to functionally characterizing miRNAs with a special emphasis on their roles in cancer. While TFs can serve either as a transcriptional activator or as a repressor, miRNAs are primarily known to confer mRNA degradation and/or translational repression by forming imperfect base-pair with the target sites primarily at the 3$'$ untranslated regions of the messenger RNAs \cite{Guo:2010kh}. While miRNAs are typically $\sim$22 nt long, several experimental studies combined with computational methods \cite{Lewis:2003uz,Doench:2004ct,Kiriakidou:2004kc,Burgler:2005ge,Lewis:2005cb,Alexiou:2009jq,Yue:2009jt} have shown that only the first six or seven consecutive nucleotides starting at the second nucleotide from the 5$'$ end of the miRNA are the most crucial determinants for target site recognition (Figure \ref{background:fig:seedmatch}). Accordingly, the 6mer or 7mer close to the 5$'$ region of the miRNA is termed as the ``seed" region or seed. miRNAs that share common seeds belong to an miRNA family as they potentially target a vastly common set of mRNAs. Moreover, the target sites at the \threeutr Watson-Crick (WC) pairing with the miRNA seed are preferentially more conserved within mammalian or among all of the vertebrate species \cite{Lewis:2003uz}. In humans, more than one third of the genes harbour sites under selective pressure to maintain their pairing to the miRNA seeds \cite{Lewis:2003uz,Grimson:2007cy}. An important variation around this seed-target pairing scheme was discovered by Lewis \emph{et al.} (2005), where the target site is flanked by a conserved adenosine `A' facing to the first nucleotide of the targeting miRNA \cite{Lewis:2005cb}.

The dynamics of the miRNA regulatory network are implicated in various phenotypic changes including embryonic development and many other complex diseases \cite{Song:2006hm,Croce:2009ff}. Although abnormal miRNA expression can sometimes be taken as a stronger indicator of carcinoma in clinical samples than aberrant mRNA expression \cite{Ramaswamy:2001hc,Lu:2005hx}, the system level mechanistic effects are usually unclear. A single miRNA can potentially target $\sim$400 distinct genes, and there are thousands of distinct endogenous miRNAs in the human genome. Thus, miRNAs are likely involved in virtually all biological processes and pathways including carcinogenesis. However, functional characterizing miRNAs hinges on the accurate identification of their mRNA targets, which has been a challenging problem due to imperfect base-pairing and condition-specific miRNA regulatory dynamics. In this section, we discuss the current state-of-art approaches in referring miRNA or miRNA-mediated transcriptional regulatory network. Table \ref{background:miRNA:methods} summarizes these methods. As we will see, each method is established through an effective unsupervised learning model by exploiting the static sequence-based information pertinent to the prior knowledge of miRNA targeting and/or the dynamic information of miRNA activities implicated by the recently available large data compendia, which interrogate genome-wide expression profiles of miRNAs and/or mRNAs across various cell conditions.

\begin{table}[!b]
  \caption{Unsupervised learning methods reviewed in this chapter}\label{background:miRNA:methods}
\begin{center}  
    \begin{tabular}{rlrr}
    \hline
    \textbf{Method} & \textbf{Algorithm} & \textbf{Section} & \textbf{Ref} \\
    \hline
    PicTar & Hidden Markov Model & \ref{background:pictar} &  \cite{Krek:2005er} \\
    TargetScore & Variational Bayesian Mixture Model & \ref{background:targetscore} & \cite{Li:2013wk}\\
    GroupMiR & Nonparametric Bayesian with Indian Buffet Process & \ref{background:groupmir} & \cite{Le:2011wx} \\
    SNMNMF & Constrained nonnegative matrix factorization & \ref{background:snmnmf} & \cite{Zhang:2011ce} \\    
    Mirsynergy & Deterministic overlapping neighbourhood expansion &  \ref{background:mirsynergy} & \cite{Li:2014fa}\\
    \hline
    \end{tabular}
\end{center}

PicTar: Probabilistic identification of combinations of Target sites; GroupMiR: Group MiRNA target prediction; SNMNMF: Sparse Network-regularized Multiple Nonnegative Matrix Factorization; PIMiM: Protein Interaction-based MicroRNA Modules.
\end{table}

\subsection{PicTar}\label{background:pictar}
PicTar (Probabilistic identification of combinations of Target sites) is one of the few models that rigorously considers the combinatorial miRNA regulations on the same target \threeutr \cite{Krek:2005er}. As an overview, PicTar first pre-filters target sites by their conservation across select species. However, the fundamental framework of PicTar is based on hidden Markov model (HMM) with a maximum likelihood (ML) approach, which is built on the logics of several earlier works from Siggia group \cite{Durbin:1998wz,Bussemaker:2000dl,Sinha:2003ui,Rajewsky:2002tt}. Among these works, PicTar was most inspired by ``Ahab", an HMM-based program developed (by the same group) to predict the combinatorial TF binding sites (TFBS) \cite{Rajewsky:2002tt}. Although PicTar has been successfully applied to three studies on vertebrates \cite{Krek:2005er} (where the original methodology paper was described), fly \cite{Grun:2005ix}, and worm \cite{Lall:2006hh} (where some improvements were described), the description of the core HMM algorithm of PicTar is rather brief. Here we will lay out the detailed technicality of the algorithm based on the information collected from several related works \cite{Durbin:1998wz,Bussemaker:2000dl,Sinha:2003ui,Rajewsky:2002tt,MBishop:2006uc}, which will help highlight its strengths, limitations, and possible future extensions.

%\begin{figure*}[!t]
%	\centering
%	\includegraphics[width=\textwidth]{figs/pictar_hmm}
%\caption[PicTar schematic view]{PicTar schematic view. Given a \threeutr and a set of $K$ miRNA sequences,  PicTar employs a $(K+1)$-state HMM to infer whether each segment of the \threeutr represents a seed match to one of the $K$ miRNAs or background (BKG). As a simple illustration, only three miRNAs (miR1, 2, 3) are shown in the HMM model on the right. }\label{background:fig:pictar}
%\end{figure*}

Let $S$ be a \threeutr sequence, $L$ the length of $S$, and $w \in \{1,\ldots,K\}$ the target sites for miRNA ``word" $w$ of length $l_w$, $p_w$ the transition probability of the occurrence of miRNA $w$, and $p_b$ the transition probability for the background of length $l_b = 1$, which is simply estimated from the fraction of A, U, G, C  (i.e., Markov model of order 0) either in $S$ with length $>$300 nt or from all query UTRs. To simplify notation, the background letters are treated as a special word $w_0$ so that $p_b\equiv p_{w_0}$ and $l_b \equiv l_{w_0} = 1$. Thus, $S$ can be represented by multiple different ways of concatenating the segments corresponding to either miRNA target sites or background. The goal is to obtain at any arbitrary nucleotide position $i$ of the 3$'$UTR sequence $S$ the posterior probability $p(\pi_i = w | S, \theta)$ that $i$ is \emph{the last position} of the word, where $\theta$ is the model parameters controlling emission probabilities (see below).

Following Markov's assumption, $p(\pi_i = w | S, \theta)$ is proportional to the products of the probabilities before and after position $i$, which can be computed in time $O(L\times K)$ by Forward-Backward algorithm as described below. Formally,
\begin{align}
	p(\pi_i = w | S, \theta) &= \frac{p(s_1, \ldots, s_L, \pi_i = w | \theta)}{p(S | \theta)}\\
	&= \frac{p(s_1, \ldots, s_i, \pi_i=w | \theta) p(s_{i+1}, \ldots, s_L | s_1, \ldots, s_i,  \pi_i=w, \theta)}{p(S | \theta)}\\
	&= \frac{p(s_1, \ldots, s_i, \pi_i=w | \theta) p(s_{i+1}, \ldots, s_L | \pi_i=w, \theta)}{p(S | \theta)} \\\label{background:posterior}
	&= \frac{Z(1,i,\pi_i=w)Z(i+1,\ldots,L|\pi_i=w)}{p(S | \theta)}
\end{align}
where $p(S | \theta)$ is the likelihood of sequence $S$ or the \emph{objective function} to be maximized in the ML framework, $Z(1,i,\pi_i=w)$ and $Z(i+1,\ldots,L|\pi_i=w)$ can be represented in recursion forms and computed via forward and backward algorithm, respectively in time $O(L\times K)$ for $K$ words. Formally, the forward algorithm is derived as follows:
\begin{align}\notag
	&Z(1, i, \pi_i=w)\\
	&= p(s_1, \ldots, s_i, \pi_i=w)\\
	&= p(s_1, \ldots, s_i | \pi_i = w) p(\pi_i=w)\\
	&= p(s_{i-l_w+1}, \ldots, s_i | \pi_i=w) p(s_1, \ldots, s_{i-l_w} | \pi_i=w) p(\pi_i=w)\\
	&= p(s_{i-l_w+1}, \ldots, s_i | \pi_i=w) p(s_1, \ldots, s_{i-l_w}, \pi_i=w)\\
	&= p(s_{i-l_w+1}, \ldots, s_i | \pi_i=w) \sum_{w'} p(s_1, \ldots, s_{i-l_w}, \pi_{i-l_w}=w') p(\pi_i=w | \pi_{i-l_w} = w')\\\label{background:pictar:fwd}
	&= e(i-l_w+1, \ldots, i | w) \sum_{w'}Z(1, i-l_w, \pi_{i-l_w}=w') p_w
\end{align}
where $e(i-l_w+1, \ldots, i | w)$ is the \emph{emission probability} assumed known (see below) and $p_w \equiv p(\pi_i=w | \pi_{i-l_w} = w')$ is the transition probability to word $w$. Note that $p_w$ is position independent. To start the recursion, $Z(1, i\le1) = 1$.

The backward algorithm is similarly derived:
\begin{align}\notag
	&Z(i+1, \ldots, L | \pi_i = w)\\
	&= p(s_{i+1}, \ldots, s_L | \pi_i = w)\\
	&= \sum_{w'} p(s_{i+1}, \ldots, s_L, \pi_{i+1}=w' | \pi_i = w)\\
	&= \sum_{w'} p(s_{i+1}, \ldots, s_L | \pi_{i+1}=w') p(\pi_{i+1}=w' | \pi_i = w)\\\notag
	&= \sum_{w'} p(s_{i+2}, \ldots, s_L | \pi_{i+1}=w') p(s_{i+2-l_{w'}}, \ldots, s_{i+1} | \pi_{i+1}=w')p(\pi_{i+1}=w' | \pi_i = w)\\\label{background:pictar:bwd}
	&= \sum_w' Z(i+2, \ldots, L | \pi_{i+1}=w') e(i+2-l_{w'}, \ldots, i+1 | w') p_{w'}
\end{align}
To start the backward recursion, $Z(L-l_w+1, L) = p_0(w)$, which is simply the frequency of word $w$ in the 3$'$UTR sequence $S$.

Given the emission probabilities, $p_w$'s (transition probability) are the only parameters that need to be set in order to maximize $p(S | \theta)$. Following the ML solution,
\begin{equation}\label{background:pictar:trans}
	p_w = \frac{\sum_ip(\pi_i = w | S, \theta)}{\sum_{w'}\sum_ip(\pi_i = w' | S, \theta)}
\end{equation}
where the posterior $p(\pi_i = w | S, \theta)$ is calculated by Eq \eqref{background:posterior}, which is in turn computed by forward-backward algorithm. Finally, the likelihood (objective) function is evaluated by a simple forward pass to the end of the sequence:
\begin{align}
	p(S|\theta) &= \sum_{w'} p(s_1, \ldots, s_L, \pi_L=w' | \theta)\\\label{background:pictar:likelihood}
	&= \sum_{w'} Z(1, L, \pi_L=w')	
\end{align}
Together, the optimization of $p_w$ in PicTar is performed using Baum-Welch algorithm for Expectation-Maximization (EM) as summarized in Algorithm \ref{background:pictarpseudo}. Finally, the PicTar score is defined as a log ratio ($F=-\log Z$) of the ML over background likelihood $F_B$:
\begin{equation}
	PicTarScore = F_B - F
\end{equation}
where $F_B$ is the likelihood when only background is considered.

%pseudocode for PicTar
\begin{algorithm}[t]
\caption{Baum-Welch HMM algorithm in PicTar \cite{Krek:2005er}}\label{background:pictarpseudo}
\begin{algorithmic}

\item Initialize $Z(1, i\le1) = 1$ and $Z(L-l_w+1, L) = p_0(w)$

\item \textbf{E-step}:

\subitem Forward recursion ($i=1,\ldots,L$): compute $Z(1, i, \pi_i=w)$ by Eq \eqref{background:pictar:fwd}

\subitem Backward recursion ($i=L-l_w,\ldots,1$): compute $Z(i+1, \ldots, L | \pi_i = w)$ by Eq \eqref{background:pictar:bwd}

\item \textbf{M-step}:

\subitem Update $p_w$ by Eq \eqref{background:pictar:trans}

\item \textbf{Evaluate likelihood}:

\subitem Compute $p(S|\theta)$ by Eq \eqref{background:pictar:likelihood}

\item Repeat \textbf{EM} steps until $p(S|\theta)$ increases by less than a cutoff

\end{algorithmic}
\end{algorithm}

As previously mentioned, the emission probabilities $e(s|w)$ in PicTar are assumed known. In Ahab for modelling TFBS, $e(s|w)$ or $m(s|w)$ is based on the position frequency matrix (PFM): $m(s|w) = \prod_{j=1}^l f_j(n|w)$, where $f_j(n|w)$ is the normalized frequency of the nucleotide $n$ at the $j\text{-}th$ position of the PFM. However, miRNAs do not have PFM. The original PicTar arbitrarily sets $e(s|w)$ to be 0.8 if there is perfect seed-match at 1-7 or 2-8 nt positions of the miRNA 5$'$ end AND the free binding energy as estimated by RNAhybrid is no less than 33\% of the optimal binding energy of the entire miRNA sequence to the UTR \cite{Lorenz:2011it}; otherwise $e(s|w)$ is set to 0.2 divided by $M$ for $M$ imperfect seed matches with only 1 mismatch allowed, provided it is above 66\% of the optimal binding energy. Thus, the setting highly disfavours imperfect seed match. The later version of PicTar changes the emission probability calculation to be the total number of occurrences in conserved \threeutr sites divided by the total number of sites  \threeutr \cite{Lall:2006hh}. The setting appears to improve the sensitive/specificity of the model but makes it more dependent on the cross-species conservation, potentially prone to false negatives (for the non-conserved but functional sites).

The major advantage of PicTar over other simpler methods is that the coordinate actions of the miRNAs (synergistic in case of optimally spaced sites or antagonistic in case of overlapping binding sites) are naturally captured within the emission and transition probabilities. For instance, the $PicTarScore$ as the joint ML of multiple miRNA target sites will be higher than the linear sum of individual miRNA target sites (i.e., synergistic effects). Longer \threeutr will score less than shorter \threeutr if both contain the same number of target sites. PicTar demonstrated a comparable signal:noise ratio relative to TargetScan and was compared favourably with some of the earlier published methods based on several surveys \cite{Alexiou:2009jq,Yue:2009jt,Sethupathy:2006df,Li:2013wk}. When applied to vertebrates, PicTar identified roughly 200 genes per miRNA, which is a rather conservative estimate compared to the recent findings by TargetScan with a conserved targeting scoring approach \cite{Friedman:2009km}. When applied to \emph{C. elegans} (worm), PicTar identified 10\% of the \emph{C. elegans} genes that are under conserved miRNA regulation.

Nonetheless, PicTar has three important limitations. First, PicTar does not consider the correlation between miRNA target sites since $p_w$ is essentially position independent. This is perhaps largely due to the increased model complexity when considering all pairwise transition probabilities between $K$ miRNAs and background since there will be $(K+1) \times K+1$ parameters to model (as apposed to only $K+1$). Supported by \cite{Grimson:2007cy}, however, the specific spatial arrangement of the target sites may be functionally important. In particular, the optimal distance between miRNA sites was estimated as 8 to $\sim$ 40 nt based on transfection followed by regression analysis \cite{Grimson:2007cy}. Although unsupported by experimental evidence, the ordering of some specific target sites may be also important. For instance, target site $x$ must be located before target site $y$ (for the same or different miRNA) to achieve optimal synergistic repression. The model that takes into account the spatial correlation between motifs is called the hcHMM (history conscious HMM) implemented in a program called Stubb for detecting TF binding sites (TFBS) rather than miRNA target predictions \cite{Sinha:2006bn}. 

Second, the ML approach is prone to local optimal especially for long UTRs or many coordinated miRNA actions considered simultaneously (i.e., many $p_w$'s). An alternative HMM formulation is to impose Bayesian priors on the HMM parameters \cite{MacKay97ensemblelearning}. In particular, \cite{Wu:2008fq} demonstrated such Bayesian formalism of HMM in modelling combinatorial TFBS. In their model so-called ``module HMM", the transition probabilities is assumed to follow a Dirichlet distribution with hyperparameters $\boldsymbol{\alpha} = \{\alpha_0, \alpha_1, \ldots, \alpha_K, \alpha_{K+1}\}$, where $\alpha_0$ corresponds to the background, $\{\alpha_1\ldots\alpha_K\}$ to the $K$ TFs, and $\alpha_{K+1}$ to the background inside of \emph{cis}-regulatory module. Inference is performed via Markov Chain Monte Carlo (MCMC) procedure or Gibbs sampling in particular. Briefly, a forward pass and backward pass are run to generate marginal probabilities at each nucleotide position. Starting at the end of the sequence, hidden states are sampled at each position based on the marginals until reaching the front of the sequence. Given the hidden states, the hyperparameters for Dirichlet distribution of the transition probabilities are then updated by simply counting the occurrences of each state. The posteriors of the hidden states at each nucleotide position are inferred as the averaged number of times the states are sampled in 1000 samplings, and the states with the \emph{maximum a posteriori} (MAP) are chosen. The model is demonstrated to perform better than Ahab and Stubb (which are the basis for PicTar) in TFBS predictions but have not yet been adapted to miRNA target predictions.

Third, since data for expression profiling of mRNAs and miRNAs by microarrays or RNA-seq is now rather abundant (e.g., \cite{Babak:2004ec}, GSE1738; GSE31568, \cite{Keller:2011jk}; GSE40499, \cite{Meunier:2012dv} from GEO) or ENCODE (GSE24565, \cite{Djebali:2013hc}) or TCGA \cite{CancerGenomeAtlasResearchNetwork:2008gr,Weinstein:2013jp}, the combinatorial regulation needs to be revisited by taking into account whether or not the co-operative miRNAs are indeed expressed \emph{in vivo} and/or the expression correlation between mRNA and miRNA. In particular, the emission probabilities $e(s|w)$ need to be redefined to integrate both sequence-based and expression-based information.

\subsection{A probabilistic approach to explore human miRNA target repertoire by integrating miRNA-overexpression data and sequence information}\label{background:targetscore}
One of the most direct way to query the targets of a given miRNA is by transfecting the miRNA into a cell and examine the expression changes of the cognate target genes \cite{Lim:2005cd}. Presumably, a \emph{bona fide} target will exhibit decreased expression upon the miRNA transfection. In particular, overexpression of miRNA coupled with expression profiling of mRNA by either microarray or RNA-seq has proved to be a promising approach \cite{Lim:2005cd,Arvey:2010ev}. Consequently, genome-wide comparison of differential gene expression holds a new promise to elucidate the global impact of a specific miRNA regulation without solely relying on evolutionary conservation. However, miRNA transfection is prone to off-target effects. For instance, overexpressing a miRNA may not only repress the expression of its direct targets but also cause a cascading repression of in-direct targets of the affected transcription activators. To improve prediction accuracy for direct miRNA targets, this chapter describes a novel model called \emph{TargetScore} that integrates expression change due to miRNA overexpression and sequence information such as context score \cite{Grimson:2007cy,Garcia:2011ec} and other orthogonal sequence-based features such as conservation \cite{Friedman:2009km} into a probabilistic score.

%\begin{figure}[!t]
%\centering\includegraphics[width=0.8\textwidth]{figs/transfection_cartoon}
%\caption[Cartoon of miRNA overexpression experiment]{Cartoon of miRNA overexpression or transfection experiment. A miRNA mimic of interest or control hairpin is transfected into a cell. True target genes are expected to exhibit expression decreases relative to the control cell.}\label{targetscore:fig:overexpression}
%\end{figure}

In one of our recent papers, we described a novel probabilistic method for miRNA target prediction problem by integrating miRNA-overexpression data and sequence-based scores from other prediction methods \cite{Li:2013wk}. Briefly, each score feature is considered as an independent observed variable, which is the input to a Variational Bayesian-Gaussian Mixture Model (VB-GMM). We chose a Bayesian over a maximum likelihood  approach to avoid overfitting. Specifically, given expression fold-change (due to miRNA transfection), we use a three-component VB-GMM to infer down-regulated targets accounting for genes with little or positive fold-change (due to off-target effects \cite{Khan:2009ki}). Otherwise,  two-component VB-GMM is applied to unsigned sequence scores. The parameters for the VB-GMM are optimized using Variational Bayesian Expectation-Maximization (VB-EM) algorithm. The mixture component with the largest absolute means of observed negative fold-change or sequence score is associated with miRNA targets and denoted as ``target component". The other components  correspond to the ``background component". It follows that inferring  miRNA-mRNA interactions is equivalent to inferring the posterior distribution of the target component given the observed variables. The targetScore is computed as the sigmoid-transformed fold-change weighted by the averaged posteriors of target components over all of the features.

%\begin{figure}[!t]
%\centering\includegraphics[width=0.9\textwidth]{figs/TargetScore}
%\caption[Bayesian Gaussian Mixture Model]{\textbf{A}. Input data consist of expression fold-change ($x_f$) due to miRNA transfection and sequence-based scores ($x_1, x_2,\ldots,x_L$). All input variables are continuous. \textbf{B}. For each score type of gene $n$ ($x_n$), we use a Variational Bayesian-Gaussian Mixture Model (VB-GMM) to infer the posterior distribution of the binary target status ($z_{nk}$), given the observed feature scores $x_n\in(x_f,x_1,x_2,\ldots x_L)$. The plate model indicates a repeating pattern of the generative model for all of the $N$ genes.}\label{targetscore:fig:vbgmm}
%\end{figure}

\subsubsection{Bayesian mixture model}\label{targetscore:bgmm}
Assuming there are $N$ genes, we denote $\mf{x} = (x_1, \ldots, x_N)^T$ as the $\log$ expression fold-change ($\mf{x}_{f}$) or sequence scores ($\mathbf{x}_{l}, l\in\{1, \ldots, L\}$). Thus,  for $L$ sets of sequence scores, $\mf{x} \in \{ \mathbf{x}_{f}, \mf{x}_1, \ldots, \mathbf{x}_{L}\}$. To simplify the following equations, we use $\mf{x}$ to represent one of the independent variables without loss of generality. To infer target genes for a miRNA given $\mf{x}$, we need to obtain the posterior distribution $p(\mathbf{z}|\mf{x})$ of the latent variable $\mf{z} \in \{z_1, \ldots, z_K\}$, where $K$=3 ($K$=2) for modelling signed (unsigned) scores such as logarithmic fold-changes (sequence scores).

We follow the standard Bayesian GMM based on \cite{MBishop:2006p485} (p474-482) with only minor modifications. Although univariate GMM ($D = 1$) is applied to each variable separately, we implemented and describe the following formalism as a more general multivariate GMM, allowing modeling the covariance matrices. Briefly, the latent variables $\mf{z}$ are sampled at probabilities $\bs{\pi}$ (mixing coefficient), that follow a Dirichlet prior $Dir(\bs{\pi} | \bs{\alpha}_0)$ with hyperparameters $\bs{\alpha}_0 = (\bs{\alpha}_{0,1}, \ldots, \bs{\alpha}_{0,K})$. To account for the relative frequency of targets and non-targets for any miRNA, we set the $\bs{\alpha}_{0,1}$ (associated with the target component) to $aN$ and other $\bs{\alpha}_{0,k} = (1 - a) \times N/(K-1)$, where $a = 0.01$ (by default). Assuming $\mf{x}$ follows a Gaussian distribution $\mathcal{N}(\mf{x}|\bs{\mu},\bs{\Lambda}^{-1})$, where $\bs{\Lambda}$ (precision matrix) is the inverse covariance matrix, $p(\bs{\mu},\bs{\Lambda})$ together follow a Gaussian-Wishart prior $\prod_k^K\mathcal{N}(\bs{\mu}_k | \mf{m}_0, (\beta_0\bs{\Lambda})^{-1})\mathcal{W}(\Lambda_k|\mf{W}_0, \nu_0)$, where the hyperparameters $\{\mf{m}_0, \beta_0, \mf{W}_0, \nu_0\} = \{\bs{\hat\mu}, 1, \mf{I}_{D\times D}, D+1\}$.

\subsubsection{Variational Bayesian Expectation Maximization}\label{targetscore:vbem}
Let $\bs{\theta} = \{\mf{z}, \bs{\pi}, \bs{\mu}, \bs{\Lambda}\}$. The marginal log likelihood can be written in terms of lower bound $\mathcal{L}(q)$ (first term) and Kullback-Leibler divergence $\mathcal{KL}(q||p)$ (second term):
\begin{equation}
	\ln p(\mf{x}) = \int q(\bs{\theta})\ln\frac{p(\mf{x},\bs{\theta})}{q(\bs{\theta})}d\bs\theta + \int q(\bs{\theta})\ln\frac{q(\bs{\theta})}{p(\bs{\theta}|\mf{x})}d\bs\theta
\end{equation}
where $q(\bs{\theta})$ is a proposed distribution for $p(\bs{\theta}|\mf{x})$, which does not have a closed form distribution. Because $\ln p(\mf{x})$ is a constant, maximizing $\mathcal{L}(q)$ implies minimizing $\mathcal{KL}(q||p)$. The general optimal solution $\ln q^*_j(\theta_j)$ is the expectation of variable $j$ \emph{w.r.t} other variables, $\mathbb{E}_{i\neq j}[\ln p(\mf{x},\bs{\theta})]$. In particular, we define $q(\mf{z}, \bs{\pi}, \bs{\mu}, \bs{\Lambda}) = q(\mf{z}) q(\bs{\pi}) q(\bs{\mu}, \bs{\Lambda})$. The expectations for the three terms (at log scale), namely $\ln q^*(\mf{z}), \ln q^*(\bs{\pi}), \ln q^*(\bs\mu,\bs\Lambda)$, have the same forms as the initial distributions due to the conjugacy of the priors. However, they require evaluation of the parameters $\{\mf{z}, \bs{\pi}, \bs{\mu}, \bs{\Lambda}\}$, which in turn all depend on the expectations of $\mf{z}$ or the posterior of interest:
\begin{equation}\label{targetscore:posterior}
	p(z_{nk}|\mf{x}_n,\bs{\theta}) \equiv \mathbb{E}[z_{nk}] = \frac{\rho_{nk}}{\sum_{j=1}^K\rho_{nj}}
\end{equation}
where $\ln \rho_{nk} = \mathbb{E}[\ln\pi_k] + \frac{1}{2}\mathbb{E}[\ln|\mf{\Lambda}_k|] - \frac{D}{2}\ln(2\pi) - \frac{1}{2}\mathbb{E}_{\mu_k,\mf{\Lambda}_k}[(\mf{x}_n - \bs{\mu}_k)^T\mf{\Lambda}_k(\mf{x}_n - \bs{\mu}_k)]$.
The inter-dependence between the expectations and model parameters falls naturally into an EM framework, namely VB-EM. Briefly, we first initialize the model parameters based on priors and randomly sample $K$ data points $\bs{\mu}$. At the $i^{th}$ iteration, we evaluate \eqref{targetscore:posterior} using the model parameters (VB-E step) and update the model parameters using \eqref{targetscore:posterior} (VB-M step). The EM iteration terminates when $\mathcal{L}(q)$ improves by less than $10^{-20}$ (default). Please refer to \cite{MBishop:2006p485} for more details.

\subsubsection{TargetScore}\label{targetscore:targetscore}
We define the targetScore as an integrative probabilistic score of a gene being a target $t$ (meaning that $z_{nk}=1$ for the target component $k$) of a miRNA:
\begin{equation}\label{targetscore:targetScore}
	\text{targetScore} = \sigma(-\log{FC}) \left(\frac{1}{L+1}\sum_{\mf{x} \in \{ \mathbf{x}_{f}, \mf{x}_1, \ldots, \mathbf{x}_{L}\}} p(t | \mf{x})\right)
\end{equation}
where $\sigma(-\log{FC}) = \frac{1}{1 + \exp(\log{FC})}$, ${\quad}p(t | \mf{x})$ is the posterior in \eqref{targetscore:posterior}.

TargetScore demonstrates superior statistical power compared to existing methods in predicting validated miRNA targets in various human cell-lines. Moreover, the confidence targets from TargetScore exhibit comparable protein downregulation and are more significantly enriched for Gene Ontology terms. Using TargetScore, we explored oncomir-oncogenes network and predicted several potential cancer-related miRNA-messenger RNA interactions. TargetScore is available at Bioconductor \url{http://www.bioconductor.org/packages/devel/bioc/ html/TargetScore.html}.

\subsection*{Network-based methods to detect miRNA regulatory modules}\label{background:nw}
Although targets of individual miRNAs are significantly enriched for certain biological processes \cite{Papadopoulos:2009kt,Tsang:2010jv}, it is also likely that multiple miRNAs are coordinated together to synergistically regulate one or more pathways \cite{Krek:2005er,Boross:2009gx,Xu:2011kl}. Indeed, despite their limited number (2578 mature miRNAs in human genome, miRBase V20, \cite{Kozomara:2014he}), miRNAs may be in charge of more evolutionarily robust and potent regulatory effects through coordinated collective actions. The hypothesis of miRNA synergism is also parsimonious or biologically plausible because the number of possible combinations of the 2578 human miRNAs is extremely large, enough to potentially react to virtually countless environmental changes. Intuitively, if a group of (miRNA) workers perform similar tasks together, then removing a single worker will not be as detrimental as assigning each worker a unique task \cite{Boross:2009gx}. 

Several related methods have been developed to study miRNA synergism. Some early methods were based on pairwise overlaps \cite{Shalgi:2007fe} or score-specific correlation \cite{Xu:2011kl} between predicted target sites of any given two (co-expressed) miRNAs. For instance, Shalgi \emph{et al.} (2007) devised an overlapping scoring scheme to account for differential 3$'$UTR lengths of the miRNA targets, which may otherwise bias the results if standard hypergeometric test was used \cite{Shalgi:2007fe}. Methods beyond pairwise overlaps have also been described. These methods considered not only the sequence-based miRNA-target site information but also the respective miRNA-mRNA expression correlation (MiMEC) across various conditions to detect miRNA regulatory modules (MiRMs).

For instance, Joung \emph{et al.} (2007) developed a probabilistic search procedure to separately sample from the mRNA and miRNA pools candidate module members with probabilities proportional to their overall frequency of being chosen as the ``fittest", which was determined by their target sites and MiMEC relative to the counterparts \cite{Joung:2007kj}. The algorithm found only the single best MiRM, which varied depending on the initial mRNA and miRNA set. Other network-based methods using either the sequence information only or using m/miRNA expression profiles only as a filter for a more disease-focused network construction on only the differentially expressed (DE) m/miRNAs. For instance, Peng \etal (2006) employed an enumeration approach to search for maximal bi-clique on DE m/miRNAs to discover complete bipartite subgraphs, where every miRNA is connected with every mRNA \cite{Peng:2009js}. The approach operated on unweighted edges only, which required discretizing miRNA-mRNA expression correlation. Also, maximal bi-clique does not necessarily imply functional MiRMs and vice versa. 

The following subsections review in details three recently developed network methods (Table \ref{background:miRNA:methods}) to detect MiRMs. Despite distinct unsupervised learning frameworks, all three methods exploit the widely available paired m/miRNA expression profiles to improve  upon the accuracy of earlier developed (sequence-based) network approaches.

\subsection{GroupMiR: inferring miRNA and mRNA group memberships with Indian Buffet Process}\label{background:groupmir}
The expression-based methods reviewed elsewhere \cite{Yue:2009jt} were essentially designed to explain the expression of each mRNA in isolation using a subset of the miRNA expression in a linear model with a fixed set of parameters. However, the same mRNAs (miRNAs) may interact with different sets of miRNAs (mRNAs) in different pathways. The exact number of pathways is unknown and may grow with an increase of size or quality of the training data. Thus, it is more natural to \emph{infer} the number of common features shared among different \emph{groups} of miRNAs and mRNAs. Accordingly, Le \emph{et al.} (2011) proposed a powerful alternative model called GroupMiR (Group MiRNA target prediction) \cite{Le:2011wx}. As an overview, GroupMiR first explored the latent binary features or memberships possessed within mRNAs, miRNAs, or shared between mRNAs and miRNAs on a potentially infinite binary feature space empowered by a \emph{nonparametric Bayesian} (NBP) formalism. Thus, the number of features was inferred rather than determined arbitrarily. Importantly, the feature assignment took into account the prior information for miRNA and mRNA targeting relationships, obtained from sequence-based target prediction tools such as TargetScan or PicTar. Based on the shared memberships, mRNAs and/or miRNAs formed groups  (or clubs). The same miRNAs (mRNAs) could possess multiple memberships and thus belong to multiple groups each corresponding to a latent feature. This was also biologically plausible since a miRNA (mRNA) may participate in several biological processes. Similar to GenMiR++ \cite{Huang:2007em}, GroupMiR then performed a Bayesian linear regression on each mRNA expression using \emph{all miRNA expression} but placing more weight on the expression of miRNAs that shared one or more common features with that mRNA.

Specifically, the framework of GroupMiR was based on a recently developed general nonparametric Bayesian prior called the Indian Buffet Process (IBP) \cite{Griffiths:2005tp} (which was later on proved to be equivalent to Beta process \cite{Thibaux:2007wn}). As the name suggests, IBP can be understood from an analogy of a type of an `Indian buffet' as follows. A finite number of $N$ customers or objects form a line to enter one after another a buffet comprised of $K$ dishes or features. Each customer $i$ samples $\sum_k\frac{m_k}{i}$ dishes selected by $m_k$ previous customers, and Poisson($\frac{\alpha}{i}$) new dishes, where $\alpha$ is a model parameter. The choices of $N$ customers on the $K$ dishes are expressed in an $N\times K$ binary matrix $\mathbf{Z}$. A left-order function $lof(\cdot)$ maps a binary matrix $\mathbf{Z}$ to a left-ordered binary matrix with columns (i.e., dishes) sorted from left to right by decreasing order of $m_k$ and breaking ties in favour of customers who enter the buffet earlier. This process defines an exchangeable distribution on \emph{equivalence class} $\mathbf{[Z]}$ comprising all of the $\mathbf{Z}$ that have the same left-ordered binary matrix $lof(\mathbf{Z})$ regardless of the order the customers enter the buffet (i.e., row order) or the dish order (i.e., column order).

Before reviewing the IBP derivation, we need to establish some notations \cite{Griffiths:2005tp}. $(z_{1k}, \ldots, z_{(i-1)k})$ ($i\in\{1,\ldots,N\}$) denotes the history $h$ of feature $k$ at object $i$, which is encoded by a single decimal number. At object $i=3$, for instance, a feature $k$ has one of four histories encoded by $0, 1, 2, 3$ corresponding to all of the four possible permutations of choices for objects 1 and 2: $(0,0), (0,1), (1,0), (1,1)$. Accordingly, for $N$ objects, there are $2^N$ histories for each feature $k$ and $2^N-1$ histories excluding the history of all zeros (i.e., $(0)_{1\times N}$). Additionally, $K_h$ denotes the number of features possessing the same history $h$, $K_0$ for all features with $m_k=0$, and $K_+=\sum_{h=1}^{2^N-1}K_h$ for all features for which $m_k>0$. Thus, $K=K_0+K_+$. It is easy to see that binary matrices belong to an equivalence class if and only if they have the same history profile $h$ for each feature $k$. The cardinality ($card$) of an equivalence class $\mathbf{[Z]}$ is the number of all of the binary matrices with the same history profile:
\begin{equation}\label{background:ibp:cardZ}
	card(\mathbf{[Z]}) = {K \choose K_0\ldots K_{2^N-1}} = \frac{K!}{\prod^{2^N-1}_{h=0}K_h!}
\end{equation}
As shown below, Eq \eqref{background:ibp:cardZ} is essential in order to establish the close-formed solution of IBP prior when $K\rightarrow\infty$ leads to an infinite feature space or infinite number of columns in $\mathbf{Z}$. After establishing the above properties, the central steps in deriving the IBP prior used in GroupMiR is reviewed below. I will focus on some steps neglected from the original work and refer the reader to the full derivation when appropriate. As in the original papers, we first derive IBP on a finite number of latent features $K$ and then take the limit making use of Eq \eqref{background:ibp:cardZ}.

Let $\mathbf{Z}$ be an $N\times K$ binary matrix, where $N = M + R$ for $M$ mRNAs and $R$ miRNAs, and $K$ is the number of latent features. Assuming the binary value $z_{k}$ in $\mathbf{Z}$ for each feature $k$ is sampled from $Bernoulli(\pi_k)$ and are conditionally independent given $\pi_k$, the joint distribution of $z_k$ is then:
\begin{align}\notag
	p(z_k|\pi_k) &= \prod_i (1-\pi_k)^{1-z_{ik}}\pi_k^{z_{ik}}\\
	&= \exp\left(\sum_i(1-z_{ik})\log(1-\pi_k) + z_{ik}\log\pi_{k}\right)
\end{align}
where $\pi_k$ follows a Beta prior $\pi_k|\alpha \sim Beta(r, s)$ with $r = \frac{\alpha}{K}, s = 1$:
\begin{equation}\label{background:ibp:pik}
	p(\pi_k|\alpha) = \frac{\pi^{r-1}_k(1-\pi_k)^{s-1}}{B(r,s)} = \frac{\pi_k^{\frac{\alpha}{k}-1}}{B(\frac{\alpha}{K}, 1)}
\end{equation}
where $B(\cdot)$ is a Beta function. To take into account the prior information between miRNA and mRNA targeting from sequence-based predictions, GroupMiR incorporated in Eq \eqref{background:ibp:pik} an $N\times N$ weight matrix $\mathbf{W}$:
\begin{equation}\label{background:groupmir:weight}
	\mathbf{W} = 
	\begin{pmatrix}
		\mathbf{0} & \mathbf{C}\\
		\mathbf{C}^T & \mathbf{0}
	\end{pmatrix}
\end{equation}
where interaction within mRNAs and within miRNAs were set to zeros and interaction between mRNA and miRNA followed the $R\times M$ scoring matrix $\mathbf{C}$ obtained from a quantitative sequence-based predictions. In particular, MiRanda scores were used in their paper. Thus, $w_{ij}$ is either 0 or defined as a pairwise potential of interactions between mRNA $i$ and miRNA $j$. The modified $p^*(\pi_k|\alpha)$ was then defined as:
\begin{equation}\label{background:ibp:pikmod}
	p^*(\pi_k|\alpha) = \frac{\pi_k^{\frac{\alpha}{K}-1}}{\mathbf{Z}'}\boldsymbol{\Phi}_{z_k}
\end{equation}
where $\boldsymbol{\Phi}_{z_k}$ and the partition function $\mathbf{Z}'$ were defined as:
\begin{align}
	\boldsymbol{\Phi}_{z_k} &= \exp\left(\sum_{i<j}w_{ij} z_{ik}z_{jk}\right)\\
	\mathbf{Z}' &= \sum_{h=0}^{2^N-1}\mathbf{\Phi}_hB(\frac{\alpha}{K}+m_h, N-m_h+1)	
\end{align}
The marginal probability of $P(\mathbf{Z})$ is derived by integrating out $\pi_k$ as follows:

\begin{align}
	P(\mathbf{Z}) &= \prod_{k=1}^K\int_{0}^{1}P(z_k|\pi_k)P(\pi_k|\alpha)d\pi_k\\
	&= \prod_{k=1}^K\int_{0}^{1}\exp\left(\sum_i (1-z_{ik})\log(1-\pi_k) + z_{ik}\log\pi_k\right)\left(\frac{\pi_k^{\frac{\alpha}{K}-1}}{\mathbf{Z}'/\boldsymbol{\Phi}_{z_k}}\right)d\pi_k\\
	&= \prod_{k=1}^K\int_{0}^{1}\frac{\boldsymbol{\Phi}_{z_k}}{\mathbf{Z}'}\exp\left(\sum_i (1-z_{ik})\log(1-\pi_k) + z_{ik}\log\pi_k\right)\exp\left[\left(\frac{\alpha}{K}-1\right)\log\pi_k\right]d\pi_k\\\label{background:ibp:mk}
	&= \prod_{k=1}^K\frac{\boldsymbol{\Phi}_{z_k}}{\mathbf{Z}'}\int_{0}^{1}\exp\left((N-m_k)\log(1-\pi_k) + m_k\log\pi_k + (\frac{\alpha}{K}-1)\log\pi_k\right)d\pi_k\\
	&= \prod_{k=1}^K\frac{\boldsymbol{\Phi}_{z_k}}{\mathbf{Z}'}\int_{0}^{1}\exp\left((N-m_k)\log(1-\pi_k)  + (\frac{\alpha}{K}+m_k-1)\log\pi_k\right)d\pi_k\\\label{background:ibp:Bfn}
	&= \prod_{k=1}^K\frac{\boldsymbol{\Phi}_{z_k}}{\mathbf{Z}'}B(\frac{\alpha}{K}+m_k, N-m_k+1)
\end{align}
where $m_k$ in Eq \eqref{background:ibp:mk} is the sum over all $z_{ik} = 1$, and  \eqref{background:ibp:Bfn} directly follows the definition of Beta function. However, when $\lim_{K\rightarrow\infty} P(\mathbf{Z}) = 0$ since the probability of sampling a specific binary matrix from an infinite number of matrices is 0. Instead, the inference was performed over the equivalence class $[\mathbf{Z}]$ with the number of $lof$-equivalent matrices defined above:
\begin{align}
	P([\mathbf{Z}]) &= \sum_{\mathbf{Z}\in[\mathbf{Z}]}P(\mathbf{Z})\\
	&= \frac{K!}{\prod^{2^N-1}_{h=0}K_h!}\prod_{k=1}^K\frac{\boldsymbol{\Phi}_{z_k}}{\mathbf{Z}'}B(\frac{\alpha}{K}+m_k, N-m_k+1) & \bigg(\text{Eq \ref{background:ibp:cardZ}, \ref{background:ibp:Bfn}}\bigg)\\\label{background:ibp:prior}
	\lim_{K\rightarrow\infty}P([\mathbf{Z}]) &= \frac{\alpha^{K_{+}}}{\prod^{2^N-1}_{h=0}K_h!}\prod_{k=1}^{K_+}\boldsymbol{\Phi}_{z_{k}}\frac{(N-m_k)!(m_k-1)!}{N!}\exp(-\alpha\boldsymbol{\Psi})
\end{align}
where
\begin{equation}
	\boldsymbol{\Psi} = \sum^{2^N-1}_{h=0}\boldsymbol{\Phi}_h\frac{(N-m_k)!(m_k-1)!}{N!}
\end{equation}
A more elaborate derivation of \eqref{background:ibp:prior} was described in the Appendix from \cite{Le:2011wx} and omitted here. Additionally, the authors also showed that when $\mathbf{W}=0$ or equivalently $\boldsymbol{\Phi}_h = 1$ for all histories $h$, then Eq \eqref{background:ibp:prior} reduces to the original IBP introduced in \cite{Griffiths:2005tp}, which is thus a special case of the weighted IBP in GroupMiR.

Given the IBP prior Eq \eqref{background:ibp:prior}, the generative process for $z_{ik}$ corresponding to an existing feature $k$ (where $m_k > 0$) was derived as follows:
\begin{align}\label{background:ibp:z}
	&P(z_{ik} =1|\mathbf{Z}_{-ik}) = \frac{P(z_{ik}=1, \mathbf{Z}_{-ik})}{P(z_{ik}=0, \mathbf{Z}_{-ik}) + P(z_{ik}=1, \mathbf{Z}_{-ik})}\\\label{background:ibp:z1}
	&= \frac{\boldsymbol{\Phi}_{z_k,z_{ik}=1}(N-m_{-ik}-1)!(m_{-ik}+1-1)!}{\boldsymbol{\Phi}_{z_k,z_{ik}=0}(N-m_{-ik})!(m_{-ik}-1)! + \boldsymbol{\Phi}_{z_k,z_{ik}=1}(N-m_{-ik}-1)!(m_{-ik}+1-1)!}\\
	\begin{split}
	&= \frac{\exp(\sum_{j\ne i}w_{ij}\cdot1\cdot z_{jk})(N-m_{-ik}-1)!(m_{-ik}+1-1)!}{\exp(\sum_{j\ne i}w_{ij}\cdot0\cdot z_{jk})(N-m_{-ik})!(m_{-ik}-1)! +}\\\label{background:ibp:z2}
	&\quad\exp(\sum_{j\ne i}w_{ij}\cdot1\cdot z_{jk})(N-m_{-ik}-1)!(m_{-ik}+1-1)!\\
	&= \frac{\exp(\sum_{j\ne i}w_{ij}z_{jk})m_{-ik}}{(N-m_{-ik})+\exp(\sum_{j\ne i}w_{ij}z_{jk})m_{-ik}}
	\end{split}
\end{align}
where the subscript ${-ik}$ (e.g., $\mathbf{Z}_{-ik}$ or $m_{-ik}$) denotes all objects for $k$ except for $i$, Eq \eqref{background:ibp:z1} arose from cancellations of the common terms in \eqref{background:ibp:prior}, and similar for \eqref{background:ibp:z2}. The number of new feature $k^*$ (where $m_{k^*}=0$) are sampled from Poisson($\frac{\alpha}{i}$).

Notably, $\mathbf{Z}$ can be expressed as $\mathbf{Z} = (\mathbf{U}^T, \mathbf{V}^T)^T$, where $\mathbf{U}$ is a $M\times K$ binary matrix for mRNA and $\mathbf{V}$ is a $R\times K$ binary matrix for miRNA. Thus, mRNA $i$ and miRNA $j$ are in the same group $k$ if $u_{ik}v_{jk}=1$. Given $\mathbf{U}$ and $\mathbf{V}$, the regression model in GroupMiR was defined as:
\begin{equation}\label{background:ibp:regression}
	x_i \sim \mathcal{N}(\mu - \sum_j(r_j + \sum_{k:u_{ik}v_{jk}=1}s_k)y_j, \sigma^2I)
\end{equation}
where $x_i$ is the expression of mRNA $i$, $\mu$ is the baseline expression for $i$, $r_j$ is the regulatory weight of miRNA $j$, $s_k$ is a group-specific coefficient for group $k$, and $y_j$ is the expression of miRNA $j$. With the Gaussian distribution assumption for $x_i$, the data likelihood then follows as:
\begin{equation}\label{background:ibp:likelihood}
	P(\mathbf{X,Y|Z},\Theta) \propto \exp\left(-\frac{1}{2\sigma^2}\sum_i(x_i - \bar{x}_i)^T(x_i - \bar{x}_i)\right)
\end{equation}
where $\Theta = (\mu, \sigma^2, \mathbf{s}, \mathbf{r})$ and $\bar{x}_i = \mu - \sum_j(r_j + \sum_{k:u_{ik}v_{jk}=1}s_k)y_j$. The (conjugate) priors over the parameters in $\Theta$ were defined and omitted here.

Finally, the marginal posterior of $\mathbf{Z}$ are defined as:
\begin{equation}\label{background:ibp:postz}
	P(z_{ik}|\mathbf{X}, \mathbf{Y}, \mathbf{Z}_{-(ik)}) \propto P(\mathbf{X,Y}|\mathbf{Z}_{-(ik)}, z_{ik})P(z_{ik}|z_{-ik})
\end{equation}
where $P(\mathbf{X,Y}|\mathbf{Z}_{-(ik)}, z_{ik})$ was obtained by integrating the likelihood (Eq \eqref{background:ibp:likelihood}) over all parameters in $\Theta$ and $P(z_{ik}|z_{-ik})$ from Eq \eqref{background:ibp:z}.

Due to the integral involved above, analytical solution for posterior in \eqref{background:ibp:postz} is difficult to obtain. Accordingly, the inference in GroupMiR was performed via MCMC:
\begin{enumerate}
	\item Sample an existing column $z_{ik}$ from Eq \eqref{background:ibp:z};
	\item Assuming object $i$ is the last customer in line (i.e., $i=N$), sample Poisson($\frac{\alpha}{N}$) new columns and sample $s_k$ from its prior (Gamma distribution) for each new column;
	\item Sample the remaining parameters by Gibbs sampler if closed-form posterior of the parameter exists (due to conjugacy) or by Metropolis-Hasting using likelihood Eq \eqref{background:ibp:likelihood} to determine the acceptance ratio;
	\item Repeat 1-3 until convergence.
\end{enumerate}
Finally, the posteriors of $\mathbf{Z}$ in Eq \eqref{background:ibp:postz} for all feature column with at least one nonzero entry serve as the target prediction of GroupMiR.

GroupMiR was applied to simulated data generated from Eq \eqref{background:ibp:regression} with $K=5$ (i.e., 5 latent features shared among miRNA and mRNA) and increasing noise level (0.1, 0.2, 0.4, 0.8) to the prior scoring matrix $\mathbf{C}$, mimicking the high false positive and negative rates from the sequence-based predictors. At the low noise levels of 0.1, 0.2, or 0.4, GroupMiR was able to identify exactly 5 latent features and above 90\% accuracy in predicting the correct memberships between miRNA and mRNA. At the high noise level of 0.8, on the other hand, GroupMiR started to identify $>5$ latent features but the accuracy remained above 90\%, demonstrating its robustness. In contrast, GenMiR++ had much a lower performance than GroupMiR on the same data, scoring lower than 60\% accuracy at the high noise level. GroupMiR was also applied to the real microarray data with 7 time points profiling the expression of miRNA and mRNA in mouse lung development. However, only the top 10\% of the genes with highest variance were chosen leading to 219 miRNAs and 1498 mRNAs. Although the authors did not justify using such a small subset of the data, it is likely due to the model complexity that prohibited the full exploration of the data. Nonetheless, GroupMiR identified higher network connectivity and higher GO enrichment for the predicted targets than GenMiR++ on the same dataset. It would have been even more convincing, however, if GroupMiR was also tested on the same datasets of 88 human tissues, which were used in the GenMiR++ study \cite{Huang:2007em}.

Although the time complexity was not analyzed for GroupMiR, it appears similar to, if not higher, than the general IBP, which has a time complexity of $O(N^{3})$ per iteration for $N$ objects \cite{DoshiVelez:2009uc}. The slow mixing rate is the main issue for IBP-based framework due to the intensive Gibbs samplings required to perform the inference, which prevented GroupMiR from fully exploring the data space at a genome scale offered by microarray or RNA-seq platforms, and consequently compromised the model accuracy. Adaptations of efficient inference algorithms that were recently developed for IBP are crucial to unleash the full power of the NPB framework \cite{DoshiVelez:2009uc}. Additionally, GroupMiR did not consider the spatial relationships between adjacent target sites of the same mRNA \threeutr for the same or different miRNAs as in PicTar (Section \ref{background:pictar}). A more biologically meaningful (IBP) prior may improve the accuracy and/or the model efficiency by  restricting the possible connections in $\mathbf{Z}$. Finally, it would be interesting to further examine the biological revelation of the groupings from $\mathbf{Z}$ on various expression consortia. In particular, miRNAs participating in many groups or having higher out-degrees in network context are likely to be more functionally important than others. Moreover, the groupings may not only reveal miRNA and mRNA targeting relationships but also the regulatory roles of mRNAs as TFs on miRNA when a modified IBP prior is used. Taken together, many directions remained unexplored with the powerful NBP framework.

\subsection{SNMNMF: sparse network-regularized multiple nonnegative matrix factorization}\label{background:snmnmf}
The nonnegative matrix factorization (NMF) algorithm was originally developed to extract latent features from images \cite{Lee:1999gw,wang2013multiple}. NMF serves as an attractive alternative to conventional dimensionality reduction techniques such as Principle Component Analysis (PCA) because it factorizes the original matrix $V_{s\times n}$ (for $s$ images and $n$ pixels) into two non-negative matrices $V_{s\times n} = W_{s\times k}H_{k\times n}$, where $W_{s\times k}$ and $H_{k\times n}$ are the  ``image encoding" and ``basis image" matrices, respectively\footnote{In the original paper \cite{Lee:1999gw}, the image matrix $V_{s\times n}$ was transposed, where the rows and columns represent the pixel and image, respectively. The representation used here is to be consistent with the one used by \cite{Zhang:2011ce} reviewed below.}. Notably, $k$ needs to be known beforehand. The non-negativity of the two factorized matrices enforced by the NMF algorithm provides the ground for intuitive interpretations of the latent features because the factorized matrices tend to be sparse and reflective to certain distinct local features of the original image matrix. Kim and Tidor (2003) were among the very first groups that introduced NMF into the world of computational biology \cite{Kim:2003hd}. In particular, the authors used NMF to assign memberships to genes based on the ``image encoding" matrix $H_{k\times n}$ in order to decipher yeast expression network using gene expression data measured by microarray. Since then, many NMF-based frameworks were developed  \cite{Devarajan:2008df}.

In particular, Zhang \emph{et al.} (2011) extended the NMF algorithm to detecting miRNA regulatory modules (MiRMs) \cite{Zhang:2011ce}. Specifically, the authors proposed a sparse network-regularized multiple NMF (SNMNMF) technique to minimize the following objective function:
\begin{align}\notag
	W, H_1, H_2 &\leftarrow \underset{W,H_1,H_2}{\arg\min} \sum_{l=1,2}||X_l - WH_l||^2 - \lambda_1Tr(H_2AH_2^T) - \lambda_2Tr(H_1BH_2^T)\\\label{background:snmnmf:obj}
	&+ \gamma_1||W||^2 + \gamma_2(\sum_j||h_j||^2 + \sum_{j'}||h_{j'}||^2)
\end{align}
where 
\begin{itemize}
	\item $X_1$ and $X_2$ are the $s\times n$ mRNA and $s\times m$ miRNA expression matrices, respectively, for $s$ samples, $n$ mRNAs, and $m$ miRNAs;
	\item $W$ is the $s\times k$ encoding matrix using $k=50$ latent features (chosen based on the number of spatially separable miRNA clusters in the human genome);
	\item $H_1$ and $H_2$ are the $k\times n$ and $k\times m$ ``image basis" matrices for genes and miRNAs, respectively;
	\item $A$ is the $n\times n$ binary gene-gene interactions matrices as a union of the transcription factor binding sites (TFBS) from TRANSFAC \cite{Wingender:2000tk} and the protein-protein interactions from \cite{Bossi:2009go};
	\item $B$ is the $n\times m$ binary miRNA-mRNA interaction matrix obtained from MicroCosm \cite{GriffithsJones:2008kb} database that hosts the target predictions from MiRanda \cite{John:2004ii};
	\item $\gamma_1||W||^2 + \gamma_2(\sum_j||h_j||^2 + \sum_{j'}||h_{j'}||^2$ are regularization terms that prevent the parameter estimates from growing too large;
	\item the weights $\lambda_{1,2}$ and regularization parameters $\gamma_{1,2}$  were selected \emph{post hoc}.
\end{itemize}
The original optimization algorithm of NMF was based on a simple gradient decent procedure, which operated on only a single input matrix. Here, however, the partial derivative of Eq \eqref{background:snmnmf:obj} with respect to each matrix ($W,H_1,H_2$) depends on the optimal solution from the other two matrices. Accordingly, the authors developed a two-stage heuristic approach, which nonetheless guarantees to converge to a local optimal: (1) update $W$ fixing $H_1, H_2$ (which are initialized randomly); (2) update $H_1,H_2$ fixing $W$, repeat 1 \& 2 until convergence. To cluster m/miRNAs, the authors exploited the encoding matrices $H_1\ ({k\times n})$ and $H_2\ ({k\times m})$ by transforming each entry into z-score: $z_{ij} = (h_{ij} - \bar{h}_{.j})/\sigma_j$, where $\bar{h}_{.j}$ and $\sigma_j$ are the mean and standard deviation of column $j$ of $H_1$ (or $H_2$) for mRNA $j$ (or miRNA $j'$), respectively. Thus, each m/miRNA was assigned to zero or more features if their corresponding z-score is above a threshold. 

SNMNMF was applied to ovarian cancer dataset containing paired m/miRNA expression profiles from TCGA measuring 559 miRNAs and 12456 genes for each of the 385 patient samples. The authors found that more than half of the 49 modules (1 module was empty) identified by SNMNMF were enriched for at least one GO terms or KEGG pathway. Also, miRNAs involved in the SNMNMF-MiRMs were enriched for cancer-related miRNAs. Moreover, Kaplan-Meier survival analysis revealed that some of the $k$ latent features from the basis matrix $W_{s\times k}$ offerred promising prognostic power. Finally, SNMNMF compared favourably with the enumeration of bi-clique (EBC) algorithm proposed by Peng \emph{et al.} (2009) in terms of the number of miRNAs involved in the modules and GO/pathway enrichments \cite{Peng:2009js}. Specifically, the authors found that EBC tended to produce modules involving only a single miRNA and multiple mRNAs. The star-shape modules are instances of a trivial case that can be derived directly from miRNA-mRNA interaction scores rather than network analysis.

Despite the statistical rigor, there are several limitations of the SNMNMF algorithm. First, the NMF approach requires a predefined number of modules in order to perform the matrix factorization, which may be data-dependent and difficult to determine beforehand. Additionally, the NMF solution is often not unique, and the identified modules do not necessarily include both miRNAs and mRNAs, which makes reproducing and interpreting the results difficult. Moreover, the SNMNMF does not enforce negative MMEC (miRNA-mRNA expression correlation), whereas the negative MMEC is necessary to ascertain the repressive function of the miRNAs on the mRNAs within the MiRMs. Finally, SNMNMF incurs a high time complexity of $O(tk(s+m+n)^2)$ for $t$ iterations, $k$ modules, $s$ samples, $m$ miRNAs, and $n$ mRNAs. Because $n$ is usually large (e.g., 12456 genes in the ovarian cancer dataset), the computation is expensive even for a small number of iterations or modules. Thus, an intuitively simple and efficient deterministic framework may serve as an attractive alternative, which we describe next.

\subsection{Mirsynergy: detecting synergistic miRNA regulatory modules by overlapping neighbourhood expansion}\label{background:mirsynergy}
In one of our recent works, we described a novel model called \emph{Mirsynergy} that integrates m/miRNA expression profiles, target site information, and gene-gene interactions (GGI) to form MiRMs, where an m/miRNA may participate in multiple MiRMs, and the module number is systematically determined given the predefined model parameters \cite{Li:2014fa}. The clustering algorithm of Mirsynergy adapts from ClusterONE \cite{Nepusz:2012cm}, which was intended to identify protein complex from PPI data. The ultimate goal here however is to  construct \emph{apriori} the MiRMs and exploit them to better explain clinical outcomes such as patient survival rate.

We formulate the construction of synergistic miRNA regulatory modules (MiRMs) as an overlapping clustering problem with two main stages. Prior to the two clustering stages, we first inferred miRNA-mRNA interaction weights (MMIW) ($\mf{W}$) using m/miRNA expression data and target site information. At stage 1, we only cluster miRNAs to greedily maximize miRNA-miRNA synergy, which is proportional to the correlation between miRNAs in terms of their MMIW. At stage 2, we fix the MiRM assignments and greedily add (remove) genes to (from) each MiRM to maximize the synergy score, which is defined as a function of the MMIW matrix and the gene-gene interaction weight (GGIW)  matrix ($\mf{H}$).

\subsubsection{Two-stage clustering}\label{mirsynergy:mm:cluster}
Let $\mf{W}$ denote the expression-based $N\times M$ MMIW matrix obtained from the coefficients of a linear regression model such as LASSO, determined as the best performing target prediction model on our data, where $w_{i,k}$ is the scoring weight for miRNA $k$ targeting mRNA $i$. Similar to the ``Meet/Min'' score defined by \cite{Shalgi:2007fe} for binary interactions of co-occurring targets of miRNA pairs, we define an $M\times M$ scoring matrix denoted as $\mf{S}$, indicating miRNA-miRNA synergistic scores between miRNA $j$ and $k$ ($j \ne k$):
\begin{equation}\label{mirsynergy:eq:mirmir}
	s_{j,k} = \frac{\sum_{i=1}^N w_{i,j}w_{i,k}}{\min[ \sum_i w_{i,j}, \sum_i w_{i,k} ]}
\end{equation}
Notably, if $\mf{W}$ were a binary matrix, Eq \ref{mirsynergy:eq:mirmir} became the ratio of number of targets shared between miRNA $j$ and $k$ over the minimum number of targets possessed by $j$ or $k$, which is essentially the original ``Meet/Min" score. We chose such scoring system to strictly reflect the overlapping between the two miRNA target repertoires rather than merely correlated trends as usually intended by alternative approaches such as Pearson correlation.

Similar to the cohesiveness defined by \cite{Nepusz:2012cm}, we define \emph{synergy} score $s(V_c)$ for any given MiRM $V_c$ as follows. Let $w^{in}(V_c)$ denote the total weights of the internal edges within the miRNA cluster, $w^{bound}(V_c)$ the total weights of the boundary edges connecting the miRNAs within $V_c$ to the miRNAs outside  $V_c$, and $\alpha(V_c)$ the penalty scores for forming cluster $V_c$. The synergy of $V_c$ (i.e., the objective function) is:
\begin{equation}\label{mirsynergy:eq:synergy}
	s(V_c) = \frac{w^{in}(V_c)}{w^{in}(V_c) + w^{bound}(V_c) + \alpha(V_c)}
\end{equation}
where $\alpha(V_c)$ reflects our limited knowledge on potential unknown targets of the added miRNA as well as the false positive targets within the cluster. Presumably, these unknown factors will  affect our decision on whether miRNA $k$ belong to cluster $V_c$. For instance, miRNA may target noncoding RNAs and seedless targets, which are the mRNAs with no perfect seed-match \cite{Helwak:2013ga}. We considered only mRNA targets with seed-match to minimize the number of false positives. By default, we set $\alpha(V_c) = 2|V_c|$, where $|V_c|$ is the cardinality of $V_c$. Additionally, we define two scoring functions to assess the overlap $\omega(V_c,V_{c'})$ between $V_c$ and $V_{c'}$ for $c\ne c'$ and the density $d_1(V_c)$ of any given $V_c$:
\begin{align}\label{mirsynergy:eq:overlap}
	&\omega(V_c,V_{c'}) = \frac{|V_c\cap V_{c'}|^2}{|V_c||V_{c'}|}\\\label{mirsynergy:eq:density}
	&d_1(V_c) = \frac{2w^{in}(V_c)}{m(m-1)}
\end{align}
where $|V_c\cap V_{c'}|$ is the total number of common elements in $V_c$ and $V_{c'}$, and $m$ is the number of miRNAs in $V_c$.

The general solution for solving an overlapping clustering problems is NP-hard \cite{barthelemy2001np}. Thus, we adapt a greedy-based approach \cite{Nepusz:2012cm}. The algorithm can be divided into two major steps. In step 1, we select as an initial \emph{seed} miRNA $k$ with the highest total weights. We then grow an MiRM $V_t$ from seed $k$ by iteratively including boundary or excluding internal miRNAs to maximize the synergy $s(V_t)$ (Eq \ref{mirsynergy:eq:synergy}) until no more node can be added or removed to improve $s(V_t)$. We then pick another miRNA that has neither been considered as seed nor included in any previously expanded $V_t$ to form $V_{t+1}$. The entire process terminates when all of the miRNAs are considered. In step 2, we treat the clusters as a graph with $V_c$ as nodes and $\omega(V_c,V_{c'}) \ge \tau$ as edges. Here $\tau$ is a free parameter. Empirically, we observed that most MiRMs are quite distinct from one another in terms of $\omega(V_c,V_{c'})$ (before the merging). Accordingly, we set $\tau$ to 0.8 to ensure merging only very similar MiRMs, which avoids producing very large MiRMs (when $\tau$ is too small). We then perform a breath-first search to find all of the weakly connected components (CC), each containing clusters that can reach directly/indirectly to one another within the CC. We merge all of the clusters in the same CC and update the synergy score accordingly.

After forming MiRMs at stage 1, we perform a similar clustering procedure by adding (removing) \emph{only the mRNAs} to (from) each MiRM. Different from stage 1, however, we grow each existing MiRM separately with no prioritized seed selection or cluster merging, which allows us to implement a parallel computation by taking advantage of the multicore processors in the modern computers. In growing/contracting each MiRM, we maximize the same synergy function (Eq \ref{mirsynergy:eq:synergy}) but changing the edge weight matrix from $\mf{S}$ to a $(N+M)\times (N+M)$ matrix by combining $\mf{W}$ (the $N\times M$ MMIW matrix) and $\mf{H}$ (the $N\times N$ GGIW matrix). Notably, here we assume miRNA-miRNA edges to be zero. Additionally, we do not add/remove miRNAs to/from the MiRM at each greedy step at this stage. Finally, we define a new density function due to the connectivity change at stage 2:
\begin{equation}\label{mirsynergy:eq:density2}
	d_2(V_c) = \frac{w^{in}(V_c)}{n(m + n - 1)}
\end{equation}
where $n$ ($m$) are the number of mRNAs (miRNAs) in the $V_c$. By default, we filter out MiRMs with $d_1(V_i) < 1e\text{-}2$ and $d_2(V_j) < 5e\text{-}3$ at stage 1 and 2, respectively. Both density thresholds were chosen based on our empirical analyses. For some datasets, in particular, we found that our greedy approach tends to produce a very large cluster involving several hundred miRNAs or several thousand mRNAs at Stage 1 or 2, respectively, which are unlikely to be biologically meaningful. Despite the ever increasing synergy (by definition), however, the anomaly modules all have very low density scores, which allows us to filter them out using the above-chosen thresholds.

Notably, standard clustering methods such as $k$-means or hierarchical clustering are not suitable for constructing MiRMs since these methods assign each data point to a unique cluster \cite{pmid23157331}. A recently developed greedy-based clustering method ClusterONE is more realistic because it allows overlap between clusters \cite{Nepusz:2012cm}. However, ClusterONE was developed with physical PPI in mind. Mirsynergy extends from ClusterONE to detecting MiRMs. The novelty of our approach resides in a two-stage clustering strategy with each stage maximizing a synergy score as a function of either the miRNA-miRNA synergistic co-regulation or miRNA-mRNA/gene-gene interactions. Several methods have incorporated GGI as PPI and TFBS into predicting MiRMs \cite{Zhang:2011ce, Le:2013hj}, which proved to be a more accurate approach than using miRNA-mRNA alone. Comparing with recent methods such as SNMNMF \cite{Zhang:2011ce} and PIMiM \cite{Le:2013hj}, however, an advantage of our deterministic formalism is the automatic determination of module number (given the predefined thresholds to merge and filter low quality clusters) and efficient computation with the theoretical bound reduced from $O(K(T+N+M)^2)$ per iteration to only $O(M(N+M))$ for $N$ ($M$) mRNA (miRNA) across $T$ samples. Because $N$ is usually much larger than $M$ and $T$, our algorithm runs orders faster. Based on our tests on a linux server, Mirsynergy took about 2 hours including the run time for LASSO to compute OV ($N$=12456; $M$=559; $T$=385), BRCA or THCA ($N$=13306; $M$=710; $T$=331 or 543, respectively), whereas SNMNMF took more than a day for each dataset. Using expression data for ovarian, breast, and thyroid cancer from TCGA, we compared Mirsynergy with internal controls and existing methods including SNMNMF reviewed above. Mirsynergy-MiRMs exhibit significantly higher functional enrichment and more coherent miRNA-mRNA expression anti-correlation. Based on the Kaplan-Meier survival analysis, we proposed several prognostically promising MiRMs and envisioned their utility in cancer research. Mirsynergy is available as an R/Bioconductor package at \url{http://www.bioconductor.org/packages/release/bioc/html/Mirsynergy.html}.

The success of our model is likely attributable to its ability to explicitly leverage two types of information at each clustering stage: (1) the miRNA-miRNA synergism based on the correlation of the inferred miRNA target score profiles from MMIW matrix; (2) the combinatorial miRNA regulatory effects on existing genetic network, implicated in the combined MMIW and GGIW matrices. We also explored other model formulations such as clustering m/miRNAs in a single clustering stage or using different MMIW matrices other than the one produced from LASSO, which tends to produce MiRMs each containing only one or a few miRNAs or several very large low quality MiRMs, which were then filtered out by the density threshold in either clustering stage. Notably, an MiRM containing only a single miRNA can be directly derived from the MMIW without any clustering approach. Moreover, Mirsynergy considers only neighbour nodes with nonzero edges. Thus, our model works the best on a sparse MMIW matrix such as the outputs from LASSO, which is the best performing expression-based methods based on our comparison with other alternatives. Nonetheless, the performance of Mirsynergy is sensitive to the quality of MMIW and GGIW. In this regard, other MMIW or GGIW matrices (generated from improved methods) can be easily incorporated into Mirsynergy as the function parameters by the users of the Bioconductor package (please refer to the package vignette for more details). In conclusion, with large amount of m/miRNA expression data becoming available, we believe that Mirsynergy will serve as a powerful tool for analyzing condition-specific miRNA regulatory networks.

%\section{Discussion}
%
%\section{Future Works}

%
%
% BibTeX users please use
%bibliographystyle{natbib}
\bibliographystyle{spmpsci}
% \bibliography{Bibliography}
 \bibliography{author}
%
% Non-BibTeX users please follow the syntax
% the syntax of "referenc.tex" for your own citations
%\input{referenc}
%%%%%%%%%%%%%%%%%%%%%%%%%%%%%%%%%%%%%%%%%%%%%%%%%%%%%%%%%%%%%%%%%%%%%%

%%%%%%%%%%%%%%%%%%%%%%%%%%%%%%%%%%%%%%%%%%%%%%%%%%%%%%%%%%%%%%%%%%%%%%

%\printindex
\end{document}